\documentclass[%
superscriptaddress,
 aps,
 amsmath,amssymb,
 noeprint,
 prl,
 twocolumn
]{revtex4-2}

\usepackage{graphicx}
\usepackage{array} 
\usepackage{xcolor}
\usepackage{dcolumn}
\usepackage[ruled,vlined]{algorithm2e}
\usepackage{braket}
\usepackage{bm}
\usepackage[colorlinks,linkcolor=blue,citecolor=red,urlcolor=blue]{hyperref}
\usepackage{times}
\usepackage{color}
\usepackage[normalem]{ulem}
\usepackage{ dsfont }
\usepackage{accents}

\newcolumntype{C}[1]{>{\centering\arraybackslash}p{#1}}

\newcommand{\ie}{i.\,e.,\ }

\newcommand{\fref}[1]{\text{Fig.}~\ref{#1}}

\newcommand{\abs}[1]{\lvert#1\rvert}

\newcommand{\tr}{\mathrm{Tr}}

\newcommand{\HarvardPhysics}{Department of Physics, Harvard University, Cambridge, MA 02138, USA}
\newcommand{\BerkeleyChemistry}{College of Chemistry, University of California Berkeley, Berkeley, CA 94720, USA}
\newcommand{\lbl}{Lawrence Berkeley National Lab, Berkeley, CA 94720, USA}
\newcommand{\HarvardSEAS}{Harvard John A. Paulson School of Engineering and Applied Sciences, Harvard University, Cambridge, MA 02138, USA}
\newcommand{\USC}{Department of Chemistry, University of Southern California, Los Angeles, CA 90089, USA}
\newcommand{\Rice}{Department of Chemistry, Rice University, Houston, TX 77005, USA}

\newcommand{\0}{0}
\newcommand{\1}{1}

\usepackage{lineno}
\usepackage{lipsum}

\begin{document}


\title{Programmable Simulations of  Molecules and Materials with Reconfigurable Quantum Processors}

 \date{\today}

\author{Nishad Maskara}
\affiliation{\HarvardPhysics}
\affiliation{\lbl}
\author{Stefan Ostermann}
\affiliation{\HarvardPhysics}
\author{James Shee}
\affiliation{\BerkeleyChemistry}
\affiliation{\Rice} 
\author{Marcin Kalinowski}
\affiliation{\HarvardPhysics}
\author{Abigail McClain Gomez}
\affiliation{\HarvardPhysics}
\author{Rodrigo Araiza Bravo}
\affiliation{\HarvardPhysics}
\author{Derek S. Wang}
\affiliation{\HarvardSEAS}
\author{Anna I. Krylov}
\affiliation{\USC}
\author{Norman Y. Yao}
\affiliation{\HarvardPhysics}
\author{Martin Head-Gordon}
\affiliation{\lbl}
\affiliation{\BerkeleyChemistry}
\author{Mikhail D. Lukin}
\affiliation{\HarvardPhysics}
\author{Susanne F. Yelin}
\affiliation{\HarvardPhysics}

\begin{abstract}
Simulations of  quantum chemistry and quantum materials are believed to be among the most important potential applications of quantum information processors, but realizing practical quantum advantage for such problems is challenging.
Here, we introduce a simulation framework for strongly correlated quantum systems
that can be represented by model spin Hamiltonians. 
Our approach leverages reconfigurable qubit architectures to programmably simulate real-time dynamics and introduces an algorithm for extracting chemically relevant spectral properties via classical co-processing of quantum measurement results.
We develop a digital-analog simulation toolbox for efficient Hamiltonian time evolution utilizing digital Floquet engineering and hardware-optimized multi-qubit operations to accurately realize complex spin-spin interactions, and as an example present an implementation proposal based on Rydberg atom arrays. 
Then, we show how detailed spectral information can be extracted from these dynamics through snapshot measurements and single-ancilla control, enabling the evaluation of excitation energies and finite-temperature susceptibilities from a single-dataset. 
To illustrate the approach, we show how this method can be used to compute key properties of a polynuclear transition-metal catalyst and 2D magnetic materials. 
\end{abstract}

\maketitle

\newpage 

\begin{figure*}
    \centering
    \includegraphics{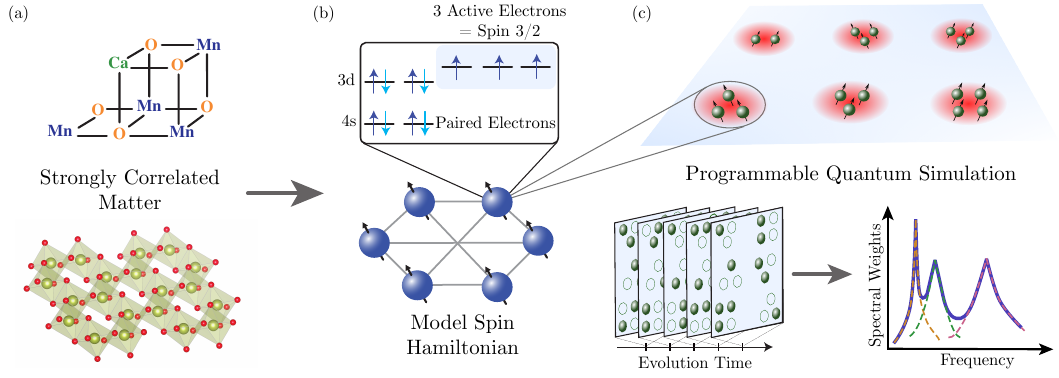}
    \caption{ \textbf{Model Hamiltonian approach to quantum simulation of strongly-correlated matter.} (a) The procedure starts with a description of the target molecule or material structure, whose electronic structure problem is reduced using classical computational chemistry techniques to a simpler effective Hamiltonian that captures the relevant low-energy behaviors.
    (b) Here we study problems which are modeled by spin Hamiltonians with potentially non-local connectivity and generic on-site spin $S \geq 1/2$, where each spin is comprised of localized, unpaired electrons in the original molecule. The key simplification comes from capturing charge fluctuations perturbatively, which is a good approximation in certain contexts.
    (c) Programmable quantum simulation is then used to calculate properties of the model Hamiltonian. We develop a simulation framework, based on encoding spins into clusters of qubits, that can be readily implemented on existing hardware.
    The toolbox enables efficient generation of complex spin interactions by leveraging dynamical reconfigurability and hardware-optimized multi-qubit gates (Figs.~\ref{fig:implementation} and~\ref{fig:ham_simulation_comparision}).
    The quantum simulator performs time-evolution under the spin Hamiltonian for various simulation times $t$, and each qubit is projectively measured to produce an set of snapshots. Subsequent classical processing extracts properties like the low-lying excitation spectrum and magnetic susceptibilities, all from the same dataset (see Figs. \ref{fig:spectroscopy_toy_model}-\ref{fig:extended_systems}).
    }
    \label{fig:overview}
\end{figure*}

A major thrust of quantum chemistry and material science involves the quantitative prediction of electronic structure properties of molecules and materials. While powerful computational techniques have been developed over the past decades, especially for weakly correlated systems~\cite{szabo2012modern,bartlett_coupled-cluster_2007,MHGDFT,BurkeDFT}, the development of tools for understanding and predicting the properties of materials that feature strongly correlated electrons remains a challenge~\cite{aoto2017arrive,hait2019levels,ahn2021designing}.
Quantum computing is a promising route to efficiently capturing such quantum correlations~\cite{alexeev_quantum_2021,bauer_quantum_2020,mcardle_quantum_2020}, and algorithms for Hamiltonian simulation and energy estimation~\cite{lin_lecture_2022} with good asymptotic scaling have been developed.
However, existing methods for simulating large-scale electronic structure problems are prohibitively expensive to run on near-term quantum hardware~\cite{beverland_assessing_2022}, highlighting the need for more efficient approaches.

One approach to capturing the complexity of strongly correlated systems utilizes model Hamiltonians~\cite{Auerbach1998}, such as the generalized Ising, Heisenberg, and Hubbard models, which describe the interactions between the active degrees of freedom at low temperatures. 
Like other coarse-graining or effective Hamiltonian approaches~\cite{bauman_quantum_2019}, model parameters can be computed from an \emph{ab initio} electronic structure problem using a number of classical techniques~\cite{bolvin_ab_2003,mayhall_computational_2014,mayhall_computational_2015,pokhilko_effective_2020,kotaru_magnetic_2023,chen_using_2022}.
Furthermore, model Hamiltonians exhibit features such as low-degree connectivity that simplify implementation, making them particularly promising candidates for quantum simulation~\cite{kandala_hardware-efficient_2017,chiesa_quantum_2019,tazhigulov_simulating_2022,wecker_solving_2015,bauer_hybrid_2016,ma_quantum_2020}.
Though approximate, simplified model Hamiltonians have proved valuable in analyzing strongly correlated problems~\cite{krewald_magnetic_2013,krewald_metal_2015,malrieu_magnetic_2014} for small system sizes, where accurate but costly classical methods can be applied.
However, as the system size increases, classical numerical methods struggle to reliably solve strongly correlated model systems, as the relevant low-energy states often exhibit a large degree of entanglement. 
In this Article, we focus on the programmable quantum simulations of \textit{spin} models. These correspond to a class of Hamiltonians that describe compounds where unpaired electrons become localized at low-temperatures and can therefore be represented as effective local spins with $S \geq 1/2$. These include many polynuclear transition metal compounds and materials containing d- and f-block elements, which play a central role in chemical catalysis and magnetism~\cite{shee2021revealing,krewald_metal_2015,pantazis_structure_2009,malrieu_magnetic_2014,calzado_role_2008,li_spin_2021,chen_using_2022,pokhilko_is_2021}. 

Recent advances in quantum simulation~\cite{monroe_programmable_2021,daley_practical_2022} have enabled the study of paradigmatic model Hamiltonians with local connectivity.
In particular, experiments have probed non-equilibrium quantum dynamics~\cite{bernien_probing_2017, keesling_quantum_2019, bluvstein_controlling_2021}, exotic forms of emergent magnetism~\cite{labuhn_tunable_2016,ebadi_quantum_2021,scholl_quantum_2021,chen_continuous_2023}, and long-range entangled topological matter~\cite{semeghini_probing_2021,satzinger_realizing_2021}, in regimes that push the limits of state-of-the-art classical simulations~\cite{kim_evidence_2023}.
The model Hamiltonians describing realistic molecules and materials however often contain more complex features, including anisotropy, non-locality, and higher-order interactions~\cite{malrieu_magnetic_2014}, demanding a higher degree of programmability~\cite{pokhilko_effective_2020}.
While universal quantum computers can in principle simulate such systems, standard implementations based on local two-qubit gates require large circuit depths~\cite{tazhigulov_simulating_2022} to realize complex interactions and long-range connectivity.
Thus, for optimal performance in devices with limited coherence, it is essential to utilize hardware-efficient capabilities to simulate such systems.

In what follows we introduce a framework to simulate model spin Hamiltonians (Fig.~\ref{fig:overview}) on reconfigurable quantum devices.
The approach combines two elements.
First, we describe a hybrid digital-analog simulation toolbox for realizing complex spin interactions, which combines the programmability of digital simulation with the efficiency of hardware-optimized multi-qubit analog operations. 
Then, we introduce an algorithm, dubbed ``many-body spectroscopy''  that leverages time-dynamics and snapshot measurements to extract detailed spectral information of the model Hamiltonian in a resource-efficient way~\cite{chan_algorithmic_2022}. 
We describe in detail how these methods can be implemented using Rydberg atom arrays~\cite{barredo_atom-by-atom_2016, cooper_alkaline-earth_2018, ma_universal_2022, singh_dual-element_2022, jenkins_ytterbium_2022}, and discuss its applicability to other emerging platforms which can support multi-qubit control and dynamic, programmable connectivity~\cite{katz_demonstration_2023}, such as the ion QCCD architecture~\cite{haffner_quantum_2008,moses2023race}.
Finally, we illustrate potential applications of the framework on model Hamiltonians describing a prototypical biochemical catalyst and 2D materials.

\subsection{Engineering spin Hamiltonians}

\begin{figure*}
    \centering
    \includegraphics{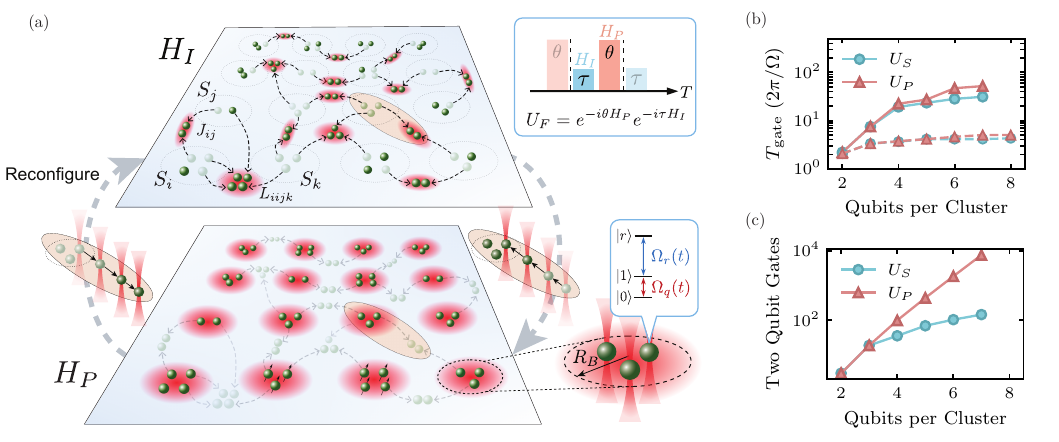}
    \caption{\textbf{Hardware-efficient implementation with neutral-atom tweezer arrays}. (a) Our protocol for programmable simulation of generic spin Hamiltonian is based on applying sequences of interactions between non-overlapping few-qubit groups. Here, we illustrate an implementation of a complex spin model using the dynamical projection approach.
    Spin-$S_i$ variables are encoded in the collective spin of a cluster of $2S_i$ qubits. Then, interactions between spins are generated by evolving pairs of qubits from each cluster under an interaction Hamiltonian $H_I$. Second-order interactions $J_{ij}$ act on two qubits, while higher-order interactions act on multiple qubits, such as the fourth-order coefficient $L_{iijk}$ of $\hat{S}_i^{\alpha} \hat{S}_i^{\beta} \hat{S}_j \hat{S}_k$.
    Then, interactions are dynamically projected into the symmetric encoding space within each cluster by evolving under Hamiltonian $H_P$. The target large spin Hamiltonian $H$ is simulated by alternately evolving under $H_I$ and $H_P$. 
    This protocol can be realized in any reconfigurable quantum processor. Here, we present an implementation for Rydberg atom arrays, which has two long-lived qubit states $\vert 0 \rangle, \vert 1 \rangle$, and an excited Rydberg state $\vert r \rangle$ with strong interactions. The interaction connectivity is dynamically changed by moving optical tweezers~\cite{bluvstein_quantum_2022}, and interactions are generated using all-to-all Rydberg blockade interactions within each cluster and simultaneous global driving of the qubit $\Omega_{q}(t)$ and Rydberg $\Omega_{r}(t)$ transitions.
    (b) Fast and efficient multi-qubit spin operations $U_S, U_P$ are identified using optimal control to optimize pulse sequences.
    Gates times are measured in units of the Rydberg driving frequency $\Omega T$, where the two-qubit CZ gate from Ref.~\cite{evered_high-fidelity_2023} takes $\Omega T / 2\pi \approx 1.2$. The alternating ansatz (solid lines) decomposes the target operations into symmetric diagonal gates and single qubit rotations, which can be are individually optimized (see Methods).  Simultaneous (dual) driving of both transitions (dotted lines) enables even faster realization of approximate $U_S$ and $U_P$ gates with ideal error rates below $\sim 10^{-3}$. 
    (c) We compare against decomposition of $U_S$ and $U_P$ into a two-qubit gate set composed of CPhase gates and single-qubit rotations. Such a decomposition rapidly becomes very costly as the cluster size grows, in contrast to the optimized hardware-native operation.
    }
    \label{fig:implementation}
\end{figure*}

The general Hamiltonian we consider is

\begin{align}
    H &= \sum_{i,\alpha} B_i^{\alpha} \hat{S}_i^{\alpha}\!+\!\sum_{ij,\alpha \beta} J_{ij}^{\alpha \beta} \hat{S}_i^{\alpha} \hat{S}_j^{\beta}\! \nonumber \\
    &+ \sum_{ijk,\alpha\beta\gamma} K_{ijk}^{\alpha \beta \gamma} \hat{S}_i^{\alpha} \hat{S}_j^{\beta} \hat{S}_k^{\gamma} + \mathrm{higher \, order},
    \label{eqn:spinH}
\end{align}

\noindent where $\hat{S}_i^{\alpha}, \alpha=x,y,z$ are spin-$S$ operators ($S\geq 1/2$) acting on the $i$-th spin, and the interaction coefficients ($J_{ij}^{\alpha\beta}$, $K_{ijk}^{\alpha\beta}$, etc.) are potentially long-range.
Hamiltonians of this form can capture the effects of many processes arising in physical compounds, including super-exchange, spin-orbit coupling, ring-exchange, and more~\cite{malrieu_magnetic_2014}.
In our approach, spin-$S$ variables are encoded into the collective spin of $2S$ qubits, such that the $i$-th spin in \eqref{eqn:spinH} is rewritten as

\begin{align}{\label{eqn:spin_to_qubit}}
\hat{S}_i^\alpha = \sum_{a=1}^{2S_i} \hat{s}_{i,a}^{\alpha},
\end{align}

\noindent where $\hat{s}_{i,a}^{\alpha}$ are the spin-1/2 operators of the $a$-th qubit in the $i$-th spin.
Valid spin-$S$ states live in the symmetric subspace with maximum total spin per site $\langle (\hat{\mathbf{S}}_i)^2 \rangle = S_i (S_i+1)$.
While several alternate approaches to encoding spins with hardware-native qudits have been proposed recently~\cite{kruckenhauser_high-dimensional_2022, chi_programmable_2022, hrmo_native_2023, gonzalez-cuadra_hardware_2022}, the cluster approach introduced here uses the same controls as qubit-based computations ensuring compatibility with existing setups~\cite{evered_high-fidelity_2023}, and naturally supports simulation of models with mixed on-site spin.

The core of our protocol involves applying a $K$ step sequential evolution under simpler interaction Hamiltonians $H_i{\;=\;}\sum_{g \in G_i} h_{i,g}$, $i\,{=}\,1,...,K$, acting on disconnected groups $G_i$ of a few qubits each.
The combined sequence realizes an effective Floquet Hamiltonian $H_F$ which approximates \eqref{eqn:spinH}.
To controllably generate effective Hamiltonians, we use the average Hamiltonian approach.
In the limit of small step-sizes $\tau \ll 1/K$, the evolution is well-approximated by $H_F^{(0)} = \frac{1}{K}\sum_{i=1}^{K} H_{i}$ to leading order, and contributions from higher-order terms are bounded~\cite{abanin_effective_2017}.

In general, the performance of this approach will be limited by simulation errors, characterized by the difference between $H_F$ and the target Hamiltonian $H$, and gate errors, determined by the hardware overhead required to implement individual evolutions $e^{-i h_{i,g} \tau}$. 
To mitigate the leading sources of error, we next develop Hamiltonian engineering protocols that leverage multi-qubit spin operations to realize \eqref{eqn:spinH} with short periodic sequences.

\subsection{Dynamical projection with digital Floquet engineering}

Our Hamiltonian engineering approach is based on the cluster encoding \eqref{eqn:spin_to_qubit}.
The key idea is to first generate interactions in the full Hilbert space using a spin-1/2 (qubit) version $H_I$ of \eqref{eqn:spinH}, and then dynamically project back onto the encoding space. 
We select $H_I$ such that projection recovers the target Hamiltonian, by mapping each $n$-body large spin interaction in \eqref{eqn:spinH} onto an equivalent $n$-body qubit interaction acting on representatives from the $n$-clusters (see Methods and Fig.~\ref{fig:implementation}a).

While this generates the target interactions between spin clusters, it also moves encoded states out of the symmetric subspace.
Therefore, to prevent evolution under $H_I$ from destroying the encoding, we alternately apply evolution under

\begin{align}{\label{eq:H_P}}
    H_P = -\lambda \sum_i P[(\hat{\mathbf{S}}_i)],
\end{align}

\noindent composed of projectors $P[(\hat{\mathbf{S}}_i)]$ onto the symmetric spin states, by applying multi-qubit gates within spin clusters. 
This Hamiltonian creates an energy gap $\lambda$ between symmetric and non-symmetric states. For $\lambda \gg J_{ij}$, the static Hamiltonian $H_I + H_P$ is effectively projected into the ground-space of $H_P$ at low-energies, realizing $H$.
However, accurate Trotter simulation in this regime, by alternating $H_I$ and $H_P$, requires two separations of time-scale between the step-size and the target interactions $\tau \ll \lambda^{-1} \ll J_{ij}^{-1}$, leading to a large number of gates.

Instead, we develop an approach which enables projection of $H_I$ onto the ground-state of $H_P$ with significantly fewer gates, using ideas inspired by dynamical decoupling~\cite{choi_robust_2020}.
This is achieved by using large-angle rotations $e^{-i \tau H_P}$, where $\tau \lambda \sim 1$, to generate a time-dependent phase on the parts of $H_I$ that couple encoded states to non-symmetric states. These phases cancel out on average, leaving the symmetric part which commutes with $H_P$.
More precisely, the combined evolution is described by

\begin{align}{\label{eqn:floquet_dyn_proj}}
    U_F &= \prod_{p=1}^{N_p} e^{-i \theta_p H_P} \prod_{j=1}^{D}e^{-i \tau H_{I,j}} \nonumber\\
    &= \prod_{p=1}^{N_p} \prod_{j=1}^{D} e^{-i \tau H_{I,j}(k)} =  e^{-i K \tau H_F},
\end{align}

\noindent where $H_I$ has been decomposed into $D$ non-overlapping groups $H_{I,j}$, $K = N_p D$ is the full length of the sequence, and $\theta_p, \tau$ parameterize the evolution times.
In the second line, we transform into an interaction picture, such that intermediate terms are evolved by $H_P$ with a cumulative phase $\Theta_k\,{=}\,\lambda^{-1}\sum_{k'<k} \theta_k$; for the rotating frame to be periodic, we require $\Theta_K \pmod {2\pi} = 0$.

Then the transformed interactions can be written as (see Methods):

\begin{align}{\label{eqn:frame_transformation}}
    H_{I,j}(k) &= e^{-i \Theta_k H_P} H_{I,j} e^{i \Theta_k H_P} \nonumber \\
    &= \underbrace{H_{I,j}^{(0)}}_{\mathrm{symmetric}} + \underbrace{ \left(\sum_{n=1}^{n_{\mathrm{max}}} e^{-i n \Theta_k} H_{I,j}^{(n)} +h.c.\right)}_{\mathrm{symmetry \; violating}}.
\end{align}

\noindent Here, $H_I^{(n)}$ captures unwanted terms that change the total spin on $n$ sites, and the symmetric part is the target Hamiltonian $H_I^{(0)}\,{=}\,H_T$.
To cancel out symmetry-violating terms at leading order in $\tau$, we choose an appropriate sequence of cumulative phases $\Theta_k\,{=}\,\lambda\sum_{k'<k} \theta_k$, such that $\sum_k e^{-i \Theta_k} = 0$.
When $n_\mathrm{max}\,{=}\,2$, \ie up to two-spin interactions, the sequence $\Theta\,{=}\,(0, 2\pi/3, 4\pi/3)$ satisfies this condition.
This sequence, being $p$-independent, produces a Floquet Hamiltonian which commutes with $H_P$ in a dressed frame~\cite{else_prethermal_2017}, and therefore preserves the encoding up to exponentially long times in $\tau^{-1}$.

This approach provides significant benefits in simulating complex spin-models, where the number of overlapping terms in $H$ grows rapidly with parameters such as the spin-size $S$, number of interactions per spin $d$, and interaction weight $n_{\mathrm{max}}$.
In this regime, a Trotter decomposition of $H$ into non-overlapping terms $h_{i,g}$ would require a sequence of length $K \geq d (2S)^{n_{\mathrm{max}}-1},$ where $d$ measures the number of interactions each spin $S_i$ is involved in.
Instead, the projection approach produces decompositions of $H$ into sequences of length $K = (n_{\mathrm{max}}+1)  \left\lceil \frac{d}{2S} \right\rceil$ (see Methods), leading to performance improvements of several orders of magnitude for models with large spins and higher-order interactions.

Similar tools can also be used realize a large class of spin circuits, by generating different evolutions $H_F^{(i)}$ during each cycle. This effectively implements discrete time-dependent evolution

\begin{align}
    U_{\mathrm{circ}} = \prod_{i=1}^{T} e^{-i \tau_i H_F^{(i)}}.
\end{align}

\noindent Variational optimization can be further used to engineer higher-order terms in $H_F$, enabling generation of more complex spin gates at no
additional cost (see Extended Data Fig. 9). Although the classical variational optimization procedure is limited to small circuits, Hamiltonian learning protocols could be used to perform larger-scale optimization on a quantum device directly~\cite{carrasco_theoretical_2021,benedetti_hardware-efficient_2021}.
Such circuits can be used for operations besides Hamiltonian simulation, including state-preparation~\cite{kandala_hardware-efficient_2017}.

\subsection{Hardware-efficient implementation}

\begin{figure}
    \centering
    \includegraphics{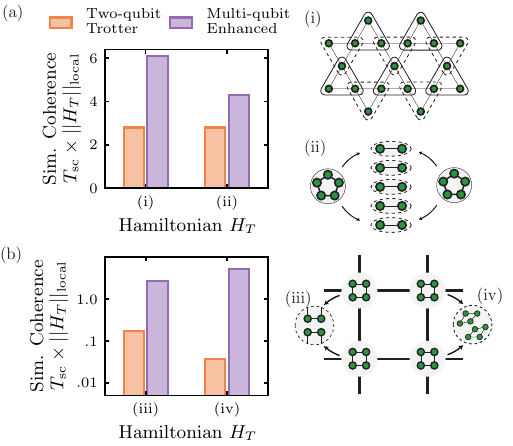}
    \caption{\textbf{Efficiency of Hamiltonian simulation framework}. 
    Estimates of the quantum simulation's coherence time $T_{\mathrm{sc}}$, in the target Hamiltonian's units $||H_{T}||_{\mathrm{local}}$ for various models.
    We consider Hamiltonian simulation implemented using the dual driving gates from Fig.~\ref{fig:implementation}b, and assume an error proportional to the gate time, such that $\Omega T/2\pi\,{=}\,1$ incurs an error of 0.1\%
    ~\cite{evered_high-fidelity_2023,scholl2023erasure,ma2023highfidelity}. 
    In all cases, we compare against an implementation using two-qubit CPhase gates with fidelity 99.9\% and perfect single-qubit rotations (see Methods for detailed descriptions of the estimation procedure).
    (a) The first two two models are (i) the spin-1/2 Kagome Heisenberg model, and (ii) two interacting spin-5/2's with Heisenberg and Dzyaloshinskii–Moriya (DM) terms, both of which are composed of only two-qubit interactions. In (i), a significant speedup is achieved by utilizing three-qubit multi-qubit gates $e^{-i \tau \mathbf{S}^2}$, which more efficiently generates Heisenberg interactions and reduces the period of the Floquet cycle from $K=4$ to $K=2$.
    In (ii) improvement is achieved using dynamical projection, which reduces $K$ from $2S$ to $2$ but at the cost of additional multi-qubit gates.
    (b) Two complex spin-models which include spin-interactions up to (iii) bi-quadratic interactions $J_1 (\mathbf{S}_i \cdot \mathbf{S}_j) + J_2 (\mathbf{S}_i \cdot \mathbf{S}_j)^2$ and (iv) bi-quartic interactions $(\mathbf{S}_i \cdot \mathbf{S}_j)^4$. These correspond to four-body and eight-body qubit interactions respectively.
    In (iii), the dramatic speedup originates from using dynamical projection to significantly reduce the Floquet period, as well as the hardware-efficiency of a native four-qubit gate. The individual contribution to the speedup from both sources is also analyzed in Methods.
    In (iv), the speedup arises fully from the hardware-efficiency of native eight-qubit operations.
    }
    \label{fig:ham_simulation_comparision}
\end{figure}

The digital Hamiltonian engineering sequence can, in principle, be realized on any universal quantum processor, but it is especially well suited for reconfigurable processors with native multi-qubit interactions~\cite{bluvstein_quantum_2022,moses2023race,katz_demonstration_2023,lis_mid-circuit_2023}.
Neutral atom arrays are a particularly promising candidate for realizing these techniques, for which we develop a detailed implementation proposal.
In this platform, two long-lived atomic states encode the qubit degree of freedom $\{\vert 0 \rangle{,}\vert 1 \rangle\}$, which can be individually manipulated with fidelity above 99.99\%~\cite{sheng_single_qubit_2018,levine_raman_2022}. Strong interactions between qubits are realized by coupling to a Rydberg state $\vert r \rangle$~\cite{jaksch_fast_2000, levine_parallel_2019}, which enables parallel multi-qubit operations~\cite{levine_parallel_2019}, with state-of-the-art two-qubit gate fidelities exceeding $99.5\%$~\cite{evered_high-fidelity_2023}.
Further, qubits can be transported with high fidelity by moving optical tweezers~\cite{bluvstein_quantum_2022}, to realize arbitrary groupings $G_i$.
By placing atoms sufficiently close together, atoms within a group can undergo strong all-to-all interactions, while interactions between groups can be made negligible by placing them far apart.

The key ingredient required for efficiently implementing $H_I$ and $H_P$ are hardware-efficient multi-qubit spin operations. We show how these can be realized by using pulse engineering to transform the native Rydberg-blockade interaction~\cite{jaksch_fast_2000, urban_observation_2009} into the desired form.
We illustrate this on two families of representative spin operations

\begin{align}{\label{eqn:target_unitaries}}
    U_S(\theta) = e^{-i \theta (\hat{\bf{S}}^2 / 2S)}, \quad U_P(\theta) = e^{-i \theta P[\hat{\bf{S}}^2]}
\end{align}

\noindent where $\hat{\bf{S}}^2$ is the total-spin operator for a cluster of $2S$ atoms, and $P[\hat{\bf{S}}^2]$ are the projectors appearing in \eqref{eq:H_P}.

One approach to engineering these operations is based on an ansatz which naturally extends Ref.~\cite{evered_high-fidelity_2023}, where $U_S$ and $U_P$ are found by optimizing an alternating sequence of diagonal phase gates and single-qubit rotations. As in prior works~\cite{jandura_time-optimal_2022, pagano_budgeting_2022, evered_high-fidelity_2023}, the pulse profiles generating symmetric diagonal operations can be obtained with numerical optimization via GrAPE~\cite{khaneja_optimal_2005}.  
For this alternating ansatz, we find a roughly linear scaling of total gate time $T_{\mathrm{gate}}$ with size of the cluster (see Fig.~\ref{fig:implementation}b).
Similar gates can also be implemented in ion-trap architectures, where coupling to collective motional modes can be used to implement diagonal phase gates~\cite{molmer_sorensen_gate_1999,katz_n_2022}.

However, Rydberg atom arrays offer additional control, which allows us to go beyond the alternating ansatz. 
Specifically, we consider simultaneously driving the qubit transition $\Omega_q(t)$ in addition to the usual Rydberg transition $\Omega_r(t)$. We find that this dual driving enables significantly faster realizations of $U_S$ and $U_P$. 
After optimizing with GrAPE to identify approximate gates with ideal fidelities above 99.9\%, 
we find total gate times below $\Omega T_{\mathrm{gate}} / 2\pi\,{=}\,6.0$ up to cluster sizes $n\,{=}\,8$ and nearly constant scaling with $n$ (see Fig.~\ref{fig:implementation}b).
These gates generate complex interactions in a very hardware-efficient way, making them ideal for accelerating spin-Hamiltonian simulations. Finally, we develop optimized decompositions of target spin operations into two-qubit gates  (see SM), and find they are still orders of magnitude more costly than the hardware-efficient implementation (Fig.~\ref{fig:implementation}c).

In Fig.~\ref{fig:ham_simulation_comparision}, we illustrate the performance of this method on four representative examples that lie within the family of Hamiltonians~\eqref{eqn:spinH}.
To quantify the performance of the simulation, we present heuristic estimates of the accessible coherent simulation time, measured in units of the target Hamiltonian's local energy scale.
We leverage access to multi-qubit spin operations of the form $e^{-i \sum_n \theta_n \mathbf{S}^n}$, and estimate gate errors based on the physical evolution time necessary to realize the target operation. 
The step-size $\tau$ is chosen to to maximize coherent simulation time, balancing simulation and gate errors (see Methods).
In the representative examples, we find the combination of dynamical projection and optimized multi-qubit gates outperforms a similarly constructed implementation based on Trotterized interactions and two-qubit gate decomposition.
Our approach significantly extends the available simulation time (up to ${\sim}$two orders of magnitude), and enables much more efficient generation of complex spin Hamiltonians.

\subsection{Spectral information from dynamics}

\begin{figure*}
    \centering
    \includegraphics{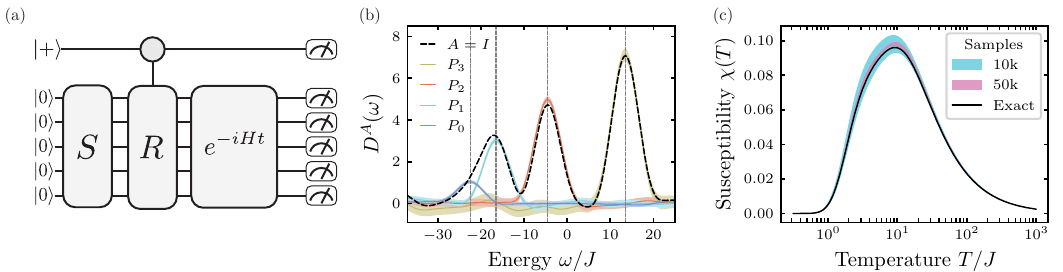}
    \caption{ \textbf{Many-body spectroscopy of model Hamiltonians}. (a) Schematic quantum circuit diagram for the algorithm. The first step is to apply a state-preparation circuit $S$ to prepare a reference state $\vert S \rangle = S \vert {0} \rangle$, followed by an ancilla-controlled perturbation $R$ preparing a superposition of $\vert S \rangle$ and the probe state $\vert R \rangle = R \vert S \rangle$. 
    This superposition is time-evolved by the target Hamiltonian, and then each qubit is projectively measured to produce a snapshot. 
    By repeating the procedure for various evolution times $t$, different perturbations $R$, and potentially different measurement bases, this setup provides access to detailed information about the spectrum of $H$.
    (b) Consider a spin Hamiltonian $H_2$ with two anti-ferromagnetically coupled spin-3/2's. We simulate 20,000 snapshot measurements and classical post processing to calculate the density-of-states $D^{\mathds{1}}(\omega)$ (black line), and total-spin resolved versions $D^{P_S}(\omega)$ (colored lines). Vertical dashed lines correspond to exact energies and colored regions represent 95\% confidence intervals.
    Peaks are broadened due to finite (coherent) simulation time $J T_{\mathrm{sim}} = 0.26$, which sets the spectral resolution. Hardware-efficient simulation schemes, which extend the simulation time (e.g. Fig.~\ref{fig:ham_simulation_comparision}), are favorable because they improve spectral resolution.
    Many-body spectroscopy further improves the effective spectral resolution, by leveraging multiple observables to distinguish overlapping peaks. Here, we see that spin-resolution significantly sparsifies the signal, enabling accurate peak detection and energy estimation, while the bare spectrum $D^{\mathds{1}}(\omega)$ is too broad to resolve all states.
    (c) The magnetic susceptibility $\chi$, can also be computed from snapshot measurements using $S^z$-resolved density-of-states (here $J T_{\mathrm{sim}} = 1.04$). 
    For these calculations, it is important to prevent exponential amplification of shot-noise. We therefore use a simple empirical truncation procedure which introduces a small amount of bias (see Methods), but enables rapid convergence with number of snapshots to the ideal value (black line). 
    }
    \label{fig:spectroscopy_toy_model}
\end{figure*}

Having described a toolbox for implementing time-evolution and state-preparation, we next present a general-purpose algorithm for calculating chemically relevant information. 
The approach, dubbed ``many-body spectroscopy'', leverages dynamical snapshot measurements and classical co-processing~\cite{chan_algorithmic_2022} to compute a wide variety of spectral quantities including low-lying states and finite-temperature properties.
The procedure, combining insights from statistical phase estimation~\cite{lin_heisenberg_limited_2022,lu_algorithms_2021,somma_quantum_2019,obrien_quantum_2019} and shadow tomography~\cite{huang_predicting_2020}, is noise-resilient and sample-efficient, making it especially promising for near-term experiments.

Specifically, the quantity we extract is an operator-resolved density of states

\begin{align}\label{eqn:raw_DOS}
    D^{A}(\omega) = \sum_n \langle n \vert A \vert n \rangle \delta(\omega - \epsilon_n),
\end{align}

\noindent where $\epsilon_n$ and $\vert n \rangle$ are the energies and eigenstates of the evolution Hamiltonian $H$, and $A$ denotes an arbitrary operator.
Spectral functions like equation~\eqref{eqn:raw_DOS} can be used to access detailed information about the properties of $H$: the location of peaks provides information about energies~\cite{lin_heisenberg_limited_2022,somma_quantum_2019,obrien_quantum_2019}, and properties of eigenstates can be computed by choosing $A$ appropriately~\cite{lu_algorithms_2021}. For example, we can compute total-spin of an eigenstate using $A = (\sum_i \hat{\textbf{S}}_i)^2$, or the local spin polarization with $A = \hat{\mathbf{S}}_i$.

The output of the quantum computation are projective measurements (snapshots), produced by the circuit depicted in Fig.~\ref{fig:spectroscopy_toy_model}a. 
First, we initialize a system of $N$ qubits and apply a state-preparation procedure to prepare a reference state $\vert S \rangle\,{=}\,S \vert {0}^N \rangle$.
Next, we use a single ancilla qubit to apply a controlled perturbation, preparing a superposition of $\vert S \rangle$ and a probe state $\vert R \rangle\,{=}\,R \vert S \rangle$,
which is then evolved under $H$ for time $t$. 
Finally each qubit is projectively measured, producing a sequence of $N+1$ bits --- a snapshot.
By measuring the ancilla in the $X$ or $Y$ basis,
this circuit effectively performs an interferometry experiment between the reference and probe states.
The resulting snapshot measurements enable parallel estimation of two-time correlation functions of the form $C_{O,R}(t) = \langle S \vert e^{i H t} O e^{-i H t} R \vert S \rangle$ for all $2^N$ operators $O$ that are diagonal in the measurement basis.

To estimate the spectral function $D^{A}(\omega)$, we use a hybrid quantum-classical computation based on the following expression,

\begin{align}{\label{eq:dos_detailed}}
    D^{A}(\omega)&=\underbrace{\tilde{\mathbb{E}}_{R\sim\mathcal{R}} \int dt}_{\mathrm{circuit\;average}}\underbrace{e^{i\omega t}\sum_{s}\langle S\vert R^{\dagger}A \, O_{s}e^{-iHt}\vert S\rangle}_{\mathrm{classical\;processing}} \nonumber\\
    &\times\underbrace{\langle S\vert e^{iHt}O_{s}e^{-iHt}R\vert S\rangle}_{\mathrm{quantum\;evolution\;(Fig.~\ref{fig:spectroscopy_toy_model})}}.
\end{align}

\noindent For this to be formally equivalent to \eqref{eqn:raw_DOS}, the distribution over perturbations $\mathcal{R}$ and ensemble of observables $\{O_s\}$ must couple uniformly to all eigenstates, to ensure unbiased estimation (see Methods). For example, given a polarized reference state $\vert 0 \rangle^{\otimes N}$, $X$-basis measurements are sufficient.

To realize the distributional average and time-integral, the quantum circuit has to be executed  (Fig.~\ref{fig:spectroscopy_toy_model}a) for randomly sampled perturbations $R$ and evolution times $t$.
In order to efficiently evaluate the classical part of \eqref{eq:dos_detailed}, we require efficient classical representations of $\vert R \rangle$ and $e^{-i H t}\vert S \rangle$.
A good choice is to select $\vert S \rangle$ to be a known eigenstate of $H$, such that the time-evolution is trivial, and preferentially sample $\vert R \rangle$ to maximize the overlap with relevant target states. 
In contrast, the quantum part of \eqref{eq:dos_detailed} includes time-evolution of $\vert R \rangle$, which has overlap with unknown eigenstates, and often includes large amounts of entanglement.
Therefore, this is estimated from snapshot measurements produced by quantum simulation (see Methods).

As an example, consider two interacting spin-3/2 particles, described by $H_2 = J \mathbf{S}_1 \cdot \mathbf{S}_2$.
We prepare a polarized reference eigenstate $\vert S \rangle = \vert 0\rangle^{\otimes 6}$, and sample perturbations $R$ from an ensemble of random single-spin rotations. This corresponds to a trivial state-preparation circuit and a simple controlled-perturbation composed of two-qubit gates in Fig.~\ref{fig:spectroscopy_toy_model}. 
Next, we measure the system in the $X$-basis, which provides access to the full spectrum for this choice of $\vert S \rangle$ (Methods).
Finally, during classical processing, the bare density of states is obtained by choosing $A = \mathds{1}$, and evaluating \eqref{eq:dos_detailed}. 
The result contains peaks at frequencies $\omega$ associated with eigenstates of $H_2$ (see Fig.~\ref{fig:spectroscopy_toy_model}b). 
Further, we can isolate individual contributions of total spin sectors by instead choosing $A=P_S$, the projectors onto $S=0,1,2,$ and $3$. This not only allows us to identify the total spin of the eigenstates, but also increases the effective spectral resolution in the presence of noise as it sparsifies the signal.
Finite-temperature response functions~\cite{alessio_equation--motion_2021}, like the $z$-component of the zero-field magnetic susceptibility $\chi(T) = \frac{1}{Z}\tr[\frac{1}{T}(S_z)^2 e^{-H/T}]$, can also be computed from the same dataset, by integrating the $S_z$-projected density-of-states $D^{S_z}(\omega)$ (see Methods).
To illustrate this, the magnetic susceptibility is extracted from the same dataset and shown in Fig.~\ref{fig:spectroscopy_toy_model}c.
The algorithm is especially promising for near-term devices, having favorable resource requirements quantified by the number of snapshots (sample complexity) and maximum evolution time (coherence) required for accurate spectral computation (see Methods for further discussion).

\begin{figure}
    \centering
    \includegraphics[width = 0.48\textwidth]{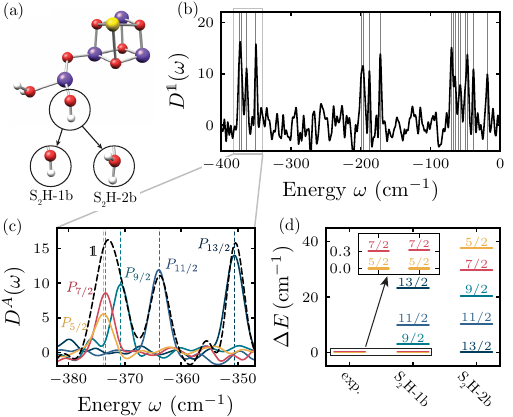}
    \caption{\textbf{Application to the oxygen-evolving complex (OEC)}.
    Our programmable quantum simulation framework can be used to compute detailed model spin Hamiltonian properties. (a) Here, we illustrate the procedure on the OEC, an organometallic catalyst with strong spin correlations. In particular, we simulate model spin Hamiltonians for two structures $\mathrm{S_2}$H-1b and $\mathrm{S_2}$H-2b, which have three spin-3/2 and one spin-2 Mn (purple) active sites (reproduced from Ref.~\cite{krewald_metal_2015} with permission from the Royal Society of Chemistry).
    Model Heisenberg coefficients for both hypothetical structures have been computed from broken-symmetry DFT~\cite{krewald_metal_2015}.
    (b) A density of states $D^{\mathds{1}}(\omega)$ calculation is simulated for the $\mathrm{S_2}$H-1b model spin Hamiltonian. Here, we use a polarized reference state $\vert{S}\rangle = \vert 0\rangle^{\otimes 13}$ a probe states $\vert R \rangle$ generated by random single-site rotations, and a evolution times $t$.
    We select 50,000 circuits with independently chosen $\vert R\rangle, t$ pairs, and draw 10 snapshots from each circuit.
    (c) Focusing on the lowest-lying states, we see three distinct peaks in $D^{\mathds{1}}(\omega)$. However, by evaluating spin-resolved quantities $D^{P_s}(\omega)$ on the same set of measurements, we identify three additional peaks, whose energies and total-spin match exact diagonalization results (vertical dotted lines).
    (d) This information is known as the spin-ladder, and can be computed using many-body spectroscopy for both the 1b and 2b states. Importantly, the spin-ladder can also be measured experimentally, and therefore can be used to help determine which structure appears in nature. In this example, experimental measurements indicate a spin-5/2 ground state and spin-7/2 first excited state. However, the ordering of low-energy states is flipped in the 2b configuration, indicating the S2H-1b hypothesis is more likely~\cite{krewald_metal_2015}.
    We note that quantities beyond total-spin can also be readily evaluated in low-lying eigenstates by inserting different operators $A$ (see Methods).
    }
    \label{fig:OEC}
\end{figure}

\subsection{Application to transition metal clusters and magnetic solids}

As an illustration of a relevant computation in chemical catalysis, we consider the Mn$_4$O$_5$Ca core of the oxygen-evolving complex (OEC), a transition metal catalyst central to photosynthesis which is still not fully understood~\cite{askerka2017o2,paul2017structural}.
Classical chemistry calculations have been used to fit model Heisenberg Hamiltonians, containing three spin-3/2 sites and one spin-2 site~\cite{krewald_magnetic_2013,krewald_metal_2015,krewald_spin_2016} (Fig.~\ref{fig:OEC}a). 
While this spin representation cannot directly capture chemical reactions, it can capture the ground and low-lying spin-states, \ie the spin-ladder, which are important in catalysis because reaction pathways depend critically on the spin multiplicity~\cite{shaik2007reactivity}.
We simulate our framework applied to the $\mathrm{S_2}$H-1b structural model from Ref.~\cite{krewald_metal_2015}, by first computing the bare density-of-states $D^{\mathds{1}}(A)$ (Fig.~\ref{fig:OEC}b).
Then, we identify a low-lying cluster of eigenstates, and compute spin-projected densities $D^{P_s}(A)$ to resolve the spin ladder (Fig.~\ref{fig:OEC}c).
The spin ladder can also be measured experimentally, providing a way to evaluate candidate models of reaction intermediates.
To highlight this, we simulate the Heisenberg model for an alternate pathway ($\mathrm{S_2}$H-2b)~\cite{krewald_metal_2015}, and observe the modification reverses the ordering of the spin ladder, indicating $\mathrm{S_2}$H-1b is more consistent with measurements (see Fig.~\ref{fig:OEC}d). 

\begin{figure}
    \centering
    \includegraphics{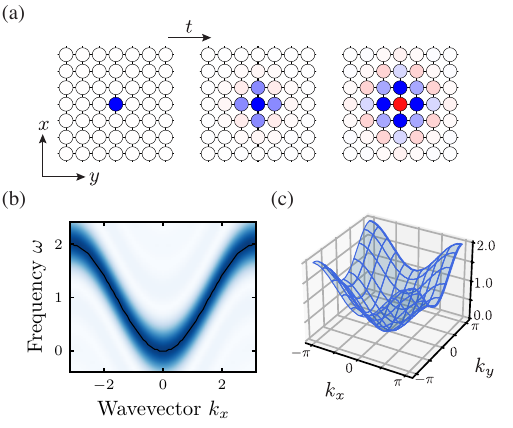}
    \caption{\textbf{Application to 2D magnetic materials.} (a) Properties of extended materials can also be investigated using our quantum simulation framework. As an example, we study the square-lattice ferromagnetic $(J>0)$ Heisenberg model $H_{\mathrm{2D}} = -J \sum_{\langle ij \rangle} \mathbf{\hat{s}}_i \cdot \mathbf{\hat{s}}_j$. By preparing the polarized ground state $\vert S \rangle = \vert 0 \rangle^{\otimes N}$, applying a single-site perturbation $R = X_\mathbf{0}$ on the central site, and measuring the system in the $X$-basis after time evolution, we can estimate the single-particle Green's function $G(\mathbf{r},t) = \langle S \vert X_\mathbf{r}(t) X_\mathbf{0} \vert S \rangle$. Therefore we select $O = X_\mathbf{r}$ for various positions $\mathbf{r}$ as the observables in~\eqref{eq:dos_detailed}, all of which are diagonal in the measurement basis. We visualize the real-part of $G(\mathbf{r},t)$, where the plotted intensity and color denotes the magnitude and sign at $Jt=0, 0.5$, and $1.0$.
    (b) The structure of excited states is extracted by classical post-processing of these measurements. Even though the spectrum is continuous, additional structure can be identified by computing the momentum-resolved density of states $D^{P_\mathbf{k}}(\omega)$, where $P_\mathbf{k}$ is a projector onto plane-wave states (see Methods).
    Restricting to $k_y=0$ and evolving to maximum time $JT_{\mathrm{max}}=8.0$, we see $D^{P_\mathbf{k}}(\omega)$ forms a band-like structure, from which a peak $\omega$ can be estimated for each $\mathbf{k}$ (black line).
    (c) This peak extraction allows us to directly estimate the single-particle dispersion $\omega(\mathbf{k})$ across the 2D Brillouin zone.}
    \label{fig:extended_systems}
\end{figure}

The framework can also be applied to study low-energy properties of extended systems, including strongly correlated materials.
We illustrate this on the ferromagnetic, square lattice Heisenberg model (Fig.~\ref{fig:extended_systems}).
For such large systems, we envision utilizing an approximate ground-state preparation method for $\vert S \rangle$, so that low-energy properties can be accessed in a noise-resiliant manner via local controlled-perturbations $R$.
Then, local Green's functions --- two-point operators at different positions and times --- can be measured to access properties of low-lying quasi-particle excitations, such as the dispersion relation of single-particle excitations (see Methods, 2D Heisenberg, for details).

\subsection{Outlook}

These considerations indicate that reconfigurable quantum processors enable a powerful, hardware-efficient framework for quantum simulation of problems from chemistry and materials science, illustrating potential directions for the search for useful quantum advantage. Specifically, in addition to the OEC, other organometallic catalysts could be studied with this approach, including iron-sulfur clusters~\cite{li_electronic_2019, tazhigulov_simulating_2022}, for which bi-quadratic terms appear in the model Hamiltonian to capture higher-order perturbative charge fluctuation effects~\cite{sharma_low-energy_2014}.
Another promising direction involves quantum simulation of low-energy properties of 2D and 3D frustrated spin systems, including model Hamiltonians for Kitaev materials~\cite{jin_unveiling_2022,xu_possible_2020,takagi_concept_2019} and molecular magnets~\cite{bogani_molecular_2008,woodruff_lanthanide_2013}.
The ability to realize non-local interactions further opens the door to simulation of spin-Hamiltonians defined on non-Euclidean interaction geometries~\cite{anshu_nlts_2023, rakovszky_physics_2023}.

The efficiency of the Hamiltonian engineering approach originates from co-designing Floquet engineering and hardware-specific multi-qubit gates. Extending this approach to larger classes of strongly-correlated model Hamiltonians is an outstanding and exciting frontier. 
In particular, it would be especially interesting to further develop the toolbox to incorporate charge transport and electron-phonon interactions. This could enable simulation of more complex model Hamiltonians, such as the t-J, Hubbard, and Hubbard-Holstein models, and expand the class of accessible chemistry problems~\cite{bauer_quantum_2020,motta_emerging_2022}.
Application to other settings, including lattice gauge theories~\cite{celi_emerging_2020, zache_fermion} and quantum optimization problems~\cite{ebadi_quantum_2022} are also of interest.
Incorporation of error mitigation and correction into the Hamiltonian simulation should be considered; specifically, the present method can potentially be generalized to control logical, encoded degrees of freedom in a hardware-efficient way~\cite{bluvstein_logical_nodate}.

Finally, characterization and development of the model Hamiltonian approach itself is an interesting and challenging problem. 
Key challenges include development of efficient schemes to compute parameters for higher-order interactions~\cite{sharma_low-energy_2014}, estimation of corrections arising from coupling to states outside the model space~\cite{gunther_more_2023}, and validation of the model Hamiltonian approximation~\cite{kotaru_magnetic_2023}. Feedback between the classical and quantum parts of the computation is an important part of these developments~\cite{bauer_hybrid_2016,vorwerk_quantum_2022,huggins_unbiasing_2022}.
For these reasons, the large-scale simulation of model Hamiltonians on quantum processors will be invaluable for testing approximations by comparing simulation outputs with experimental measurements.
Hence, the approach proposed in this work facilitates exciting directions in computational chemistry and quantum simulation, aiming towards constructing a novel \emph{ab initio} simulation pipeline that utilizes hybrid quantum-classical resources.

\bibliography{QuantumChemistry}

\noindent{\bf Acknowledgments.} We thank A. Aldossary, T. Betley, D. Bluvstein, M. Cain, L. Cunha, J. Feldmeier, C. Kokail, J. Lee, N. Leitao, S. Gant, J. Haber, S. Hollerith, T. Manovitz, J. Neaton, S. Sachdev, A. Schuckert, K. Seethram, and P. Zoller for insightful discussions.  
This work was supported by the US Department of Energy [DE-SC0021013 and DOE Quantum Systems Accelerator Center (contract no. 7568717)], the Defense Advanced Research Projects Agency (grant no. W911NF2010021), the National Science Foundation (via CUA PFC, Q-IDEAS HDR, PHY-2207972), the Department of Defense Multidisciplinary University Research Initiative (ARO MURI, grant no. W911NF2010082), Wellcome Leap under the Q for Bio program, the DARPA IMPAQT programme (grant number HR0011-23-3-0030) and the Harvard-MIT Center for Ultracold Atoms. N.M. acknowledges support from the Department of Energy Computational Science Graduate Fellowship under Award Number DE-SC0021110. 
A.M.G and R.A.B acknowledges support from the NSF through the Graduate Research Fellowships Program, and A.M.G also acknowledges support through the Theodore H. Ashford Fellowships in the Sciences. 

\resetlinenumber
\section{Appendix}

\subsection{Hamiltonian Engineering}

The Hamiltonian engineering toolbox introduced here is based on the average Hamiltonian approach. This approach uses the fact that in the high-frequency limit, the effective Floquet cycle period $K \tau$ is much smaller than the inverse local energy scales of $H(k)$, and the Floquet Hamiltonian $H_F$ can be well-approximated by expanding in a small parameter $\frac{K\tau}{||H(k)||_{\mathrm{local}}}$~\cite{abanin_effective_2017}.
The leading contribution is the \textit{average Hamiltonian}

\begin{align}
    H_{F}^{(0)} = \frac{1}{K}\sum_{k=0}^{K-1} H(k).
\end{align}

\noindent The second order term also takes a simple form, 

\begin{align}
    H_{F}^{(1)} = \frac{\tau}{2K} \sum_{k<k'} \left[ H(k),H(k') \right].
\end{align}

\noindent The results presented in the main text involve engineering the average Hamiltonian $H_F^{(0)}$ to reproduce the target \eqref{eqn:spinH}.
In this setting, the second order term generates simulation errors, and is order $O(\tau)$.
It can be cancelled by selecting time-reversal symmetric sequences of length $2K$, where the second half of the pulse is defined by $\Theta_k = \Theta_{K-1-k}$.
This reduces simulation errors to order $O(\tau^2)$, but might potentially alter the pre-thermal properties of the Floquet Hamiltonian, which is an interesting problem for further research.

Going beyond average Hamiltonian engineering is also possible by optimizing the Floquet sequence~\cite{kalinowski_non-abelian_2023,benedetti_hardware-efficient_2021,tepaske_optimal_2023}.
We demonstrate engineering of higher-order terms for a small system composed of two interacting spin-3/2's, to controllably engineer up to bi-cubic terms $(\hat{\mathbf{S}}_i \cdot \hat{\mathbf{S}}_j)^3$ using only two- and three-qubit operations in  Extended Data Fig.~\ref{fig:supp_gates_and_higher_order} and SM.

\subsection{High-spin Hamiltonian engineering with dynamical Floquet projection}

To compute the form of the interaction Hamiltonian in the rotated frame $H_I(k) = e^{i \Theta_k H_P} H_I e^{-i \Theta_k H_P}$ as used in \eqref{eqn:frame_transformation} let's consider a single spin-1/2 particle belonging to a spin-$S_i$ cluster, and define projectors $P_i = P[\hat{\bf{S}}_i^2]$ onto the symmetric space and $Q_i = \mathds{1} - P_i$ onto its complement.
The spin-1/2 term can be split into four parts

\begin{align}
    \hat{\bf{s}}_{i,1} &= \underbrace{P_i \hat{\bf{s}}_{i,1} P_i + Q_i \hat{\bf{s}}_{i,1} Q_i}_{\rm{symmetric}} + \underbrace{P_i \hat{\bf{s}}_{i,1} Q_i + Q_i \hat{\bf{s}}_{i,1} P_i}_{\rm{non-symmetric}}
\end{align}

\noindent where the first two terms preserve the on-site total spin, and the second two change the on-site total spin.
Since $H_P$ acts as $1-P_i$ on the $i$-th spin, we label terms by how they change the expectation value of $(1-P_i)$

\begin{align}
    \hat{\bf{s}}_{i,1}^{(0)} &= P_i \hat{\bf{s}}_{i,1} P_i + Q_i \hat{\bf{s}}_{i,1} Q_i \\
    \hat{\bf{s}}_{i,1}^{(+1)}&= Q_i \hat{\bf{s}}_{i,1} P_i\\
    \hat{\bf{s}}_{i,1}^{(-1)}&= P_i \hat{\bf{s}}_{i,1} Q_i
\end{align}

This rule can be extended to higher-weight operators. For example, two-spin interactions between clusters $h_{ij} = \hat{\bf{s}}_{i,1} \cdot \hat{\bf{s}}_{j,1}$ decompose into five parts $h_{ij}^{(n)}, n=-2,-1,0,1,2$,

\begin{align}
    h_{ij}(k) &= \underbrace{\hat{\bf{s}}_{i,1}^{(0)} \cdot \hat{\bf{s}}_{j,1}^{(0)} + \hat{\bf{s}}_{i,1}^{(+1)} \cdot \hat{\bf{s}}_{j,1}^{(-1)} + \hat{\bf{s}}_{i,1}^{(+1)} \cdot \hat{\bf{s}}_{j,1}^{(-1)}}_{h_{ij}^{(0)}} \\
    &+ \underbrace{\left(\hat{\bf{s}}_{i,1}^{(+1)} \cdot \hat{\bf{s}}_{j,1}^{(0)} + \hat{\bf{s}}_{i,1}^{(0)} \cdot \hat{\bf{s}}_{j,1}^{(+1)} \right)}_{h_{ij}^{(+1)}} + h.c. \\
    &+ \underbrace{\left(\hat{\bf{s}}_{i,1}^{(+1)} \cdot \hat{\bf{s}}_{j,1}^{(+1)} \right)}_{h_{ij}^{(+2)}} + h.c.,
\end{align}

\noindent corresponding to the different ways to change the expectation value of $2\,{-}\,P_i\,{-}\,P_j$.
$n$-spin interactions will have parts running from $h^{(-n)}$ to $h^{(+n)}$.
Therefore, in the rotating frame, the Hamiltonian terms transform as 
\begin{align}
    h(k) = \sum_{n=-n_{\mathrm{max}}}^{n_{\mathrm{max}}} h^{(n)} e^{i \Theta_k n}.
\end{align}

We further choose a sequence of $\Theta_k$'s such that only the $h^{(0)}$ contribution is non-zero on average.
The simplest sequence which satisfies these conditions are a family of cyclic pulses of order $P$.

\begin{align}
    \Theta_{k}&=\frac{2\pi i}{P} k & k=0,...,P-1
\end{align}

\noindent which satisfy the cancellation condition as long as $n_{\mathrm{max}} < P$.
In the two-body case, of the terms which contribute to $h_{ij}^{(0)}$, only $\hat{\bf{s}}_{i,1}^{(0)} \cdot \hat{\bf{s}}_{j,1}^{(0)}$ acts non-trivially in the symmetric subspace, so we focus on this term.
An explicit form can be computed by decomposing $\hat{\bf{s}}_{i,1}$ into its permutation-symmetric and orthogonal components,

\begin{align}{\label{eqn:spinhalf_to_sym}}
    \hat{\bf{s}}_{i,1} &= \underbrace{\frac{1}{2S_i} \sum_{a=1}^{2S_i} \hat{\bf{s}}_{i,a}}_{\rm{symmetric}} + \underbrace{\left(\frac{2S_i-1}{2S_i}\hat{\bf{s}}_{i,1} - \frac{1}{2S_i}\sum_{a'=2}^{2S_i} \hat{\bf{s}}_{i,a'} \right)}_{\rm{non-symmetric}}.
\end{align}

\noindent Therefore, the symmetric part $\hat{\bf{s}}_{i,1}^{(0)} = \frac{1}{2S_i} \hat{\bf{S}}_i$ is proportional to the collective spin.

With this understanding, we can construct a spin-1/2 interaction Hamiltonian $H_I$ that recovers \eqref{eqn:spinH} under projection, by replacing each $n$-site high-spin interaction with an analagous spin-1/2 one. For example, for $n=2$ the replacement proceeds as

\begin{align}
    J_{ij}^{\alpha \beta} \hat{S}_i^{\alpha} \hat{S}_j^{\beta} &\rightarrow \overline{J_{ij}}^{\alpha \beta} \hat{s}_{i,a}^{\alpha} \hat{s}_{j,b}^{\beta} \nonumber\\
    \overline{J_{ij}}^{\alpha \beta} &= 4 S_i S_j J_{ij}^{\alpha \beta},
\end{align}

\noindent where intra-cluster indexes $a,b$ encode which representative from spin's $i$ and $j$ are used to generate the interaction.
The interaction strength is further boosted to $\overline{J_{ij}}$, to account for the $\frac{1}{2S}$ factor in ~\eqref{eqn:spinhalf_to_sym}. 
A straightforward calculation shows that for higher-weight interactions (e.g. $n_{\mathrm{max}}$), the interaction should also be boosted (e.g. $\overline{K_{ijk}} = 8 S_i S_j S_k K_{ijk}$) to recover the target large-spin operator under projection.

In general, it may not be feasible to  uniquely assign each spin-$S_i$ interaction in $H$ to qubits in $H_I$ especially when a spin-$S_i$ is involved in more than $2S$ interactions. In this case, $H_I$ can be implemented by splitting it into a sequence of non-overlapping groups $H_{I,1},...,H_{I,D}$ that approximate $H_I$ on average. Each sequence can handle up to $D(2S)$ interactions per spin, so if $d$ is the interaction degree then we require a sequence of length $D = \lceil \frac{d}{2S} \rceil$.  While manual decompositions sufficed for the models studied here, automated methods to determine efficient decompositions will have to be developed for more complex systems.

\subsection{Multi-qubit gates with Rydberg blockade}

The Rydberg Hamiltonian governing a cluster of $N$ atoms is

\begin{align}\label{eqn:Ryd_Ham}
    H_{\mathrm{cluster}} &= \frac{\Omega_q(t)}{2} \sum_i  \vert 1 \rangle_i \langle 0 \vert + \frac{\Omega_r(t)}{2} \sum_i \vert r \rangle_i \langle 1 \vert + h.c. \nonumber\\
    &+ \sum_{i<j} V_{ij} \vert r \rangle_i \langle r \vert \otimes \vert r \rangle_j \langle r \vert
\end{align}

\noindent where $\Omega_q(t), \Omega_r(t)$ are complex valued driving fields.
In the blockade approximation, which is valid when $V_{ij} \gg \Omega_r$, there is at most one atom in state $\vert r \rangle$. 
Therefore, at leading order in $\Omega_r / V_{ij}$, $H_{\mathrm{cluster}}$ is approximated by projecting into the manifold of blockade consistent states.
This produces an interacting model with an emergent permutation symmetry (see SM). This symmetry allows us to write $H_\mathrm{cluster}$ in a low-dimensional representation of the Hilbert space scaling as $O(N N_S)$ for a representation including the $N_S$ largest total-spin sectors (see SM).

Figure~\ref{fig:implementation}b shows optimization results for an alternating ansatz with separate $\Omega_r(t)$ and $\Omega_q(t)$ applications (solid lines) versus a dual driving scheme with simultaneous field control (dashed lines). For the alternating ansatz the optimization process begins with finding short sequences of symmetric diagonal gates $D(\phi)$ and global single-qubit rotations $Q(\theta)$ that combined realize $U_S$ and $U_P$ (see SM). While $Q(\theta)$ uses global $\Omega_q(t)$ control, $D(\phi)$ involves multi-qubit interactions, and the pulse sequences to realize $D(\phi)$ can be optimized through gradient ascent pulse engineering (GrAPE)~\cite{khaneja_optimal_2005,jandura_time-optimal_2022, evered_high-fidelity_2023}. Numerical optimization is feasible due to the  manageable size of the low-dimensional basis. We find that the maximum gate time $T^*(n)$ for generic phases (see SM).
Gate times in Fig.~\ref{fig:implementation} are computed by multiplying $T^*(n)$ with the shortest sequence length determined in the first step.
The operations $Q(\theta)$ and $D(\phi)$ can also be promoted to controlled operations (see SM), as required for the controlled perturbation in Fig.~\ref{fig:spectroscopy_toy_model}a. 

Dual driving gate profiles are directly optimized using GrAPE, with an added smoothness regularization to ensure driving profiles can be implemented with available classical controls (see SM and Extended data Fig.~\ref{fig:supp_gates_and_higher_order}).
Interestingly the optimized fidelity is roughly system size independent, up to error rates around $10^{-3}$. We select this fidelity threshold, and plot the resulting gate times in Fig.~\ref{fig:implementation}b (dotted lines).
For more stringent thresholds, such as $10^{-6}$, we recover an approximately linear dependence in $n$, producing gate times comparable to the alternating decomposition.

For comparison, we estimate gate counts for a decomposition of $U_S$ and $U_P$ into single and two qubit gates (see Fig.~\ref{fig:implementation}c).
Using the Qiskit transpiler~\cite{Qiskit}, we find two-qubit decompositions into single qubit rotations and CPhase gates for $U_P$.
For $U_S$, the diagonal gates $D(\phi)$ found numerically in the alternating ansatz only require two-qubit interactions, and can be decomposed into ${n \choose 2}$ two-qubit CPhase gates, outperforming the Qiskit result for $n \geq 3$.
Finally, we show in SM that symmetric operations like $U_S$ and $U_P$ can be implemented with $\mathrm{poly}(n)$ two-qubit gates using ancilla qubits, by constructing an efficient MPO representation of arbitrary $n$-spin operations.

The fidelity of the multi-qubit gates is subject to errors like spontaneous emission and dephasing due to a finite $T_2^*$, as seen in current Rydberg gates~\cite{evered_high-fidelity_2023}. These errors can be mitigated by improving the excitation schemes. Specifically, single-photon schemes as used in as in Refs.~\cite{ma2023highfidelity,scholl2023erasure}, avoiding intermediate-state scattering, may enhance performance for bigger clusters. This is because the rate of decay from the Rydberg state does not depend on cluster size, since the number of Rydberg excitations is never larger than one.

\subsection{Estimating simulation time}

The simulation time is defined as the maximum evolution time, before which the typical error-per-qubit is below some target threshold $\epsilon$.
We account for both coherent Hamiltonian simulation errors and incoherent gate errors.
For a symmetrized sequence, we estimate the scaling of both contributions to be 

\begin{align}
    \varepsilon_{\mathrm{sim}} = (c_2 \tau^2)^2 T^2, \varepsilon_{\mathrm{gate}} = \frac{g T}{\tau},
\end{align}

\noindent with target evolution time $T$ and step-size $\tau$.
Here, the coefficient $c_2$ depends on the detail of the Hamiltonian simulation protocol, and can be estimated from numerics or the third order term in the Magnus expansion (see SM).
The coefficient $g$ measures the gate error probability per cycle, and is determined by asuming each multi-qubit gate has a fixed probability of failure $T_{\mathrm{gate}} g_0$ which scales linearly in the time of the gate.
For simplicity, we utilize the estimated $T_{\mathrm{gate}}$ for large-angle unitary $U_P$ with $\theta = \pi$.
Both $c_2$ and $g$ are estimated to grow extensively in system size. Thus, we work instead with the intensive version of these quantities, $\tilde{c}_2 = c_2/L$ and $\tilde{g} = g/L$.

Optimizing the step-size $\tau$ to minimize error (see SM), the maximum evolution time scales as 

\begin{align}
    T_{\mathrm{opt}} = \frac{2^{2/3}}{5^{5/6}} \frac{\varepsilon^{5/6} / L}{(\tilde{c}_2 \tilde{g}^2)^{1/3}}
\end{align}

\noindent for a target error rate $\varepsilon$.

In Fig.~\ref{fig:ham_simulation_comparision}, we use this formula, along with numerical estimates for $c_2$ in models (i) and (ii), and heuristic estimates for models (iii) and (iv).
We further estimate $g$ using the simultaneous driving gate times of Fig.~\ref{fig:implementation}b, and select an error per Rabi cycle of $g_0 = 10^{-3}$.
The target error we select is $\varepsilon^{5/6} / L = 0.1$, which grows approximately extensively with system size.

We illustrate the benefits of our approach on four example Hamiltonians (Fig.~\ref{fig:ham_simulation_comparision}).
First we consider the spin-1/2 Heisenberg model on a Kagome lattice (i).
The standard two-qubit decomposition involves applying a sequence of six-steps, each applying a $e^{-i \theta \mathbf{s}_i \cdot \mathbf{s}_j}$ gate along an edge $(i,j)$ (solid lines in Fig.~\ref{fig:ham_simulation_comparision}).
However, using native three-qubit Heisenberg gates, we group the interactions into upwards and downwards facing triangles (dotted lines Fig.~\ref{fig:ham_simulation_comparision}), reducing the sequence to only two-steps, improving $c_2$. The three-qubit gate also generates interactions more efficiently, further improving $g$. Note that four-qubit version of this scheme could simulate the spin-1/2 Heisenberg model on a pyrochlore lattice using $K=2$~\cite{hermele_pyrochlore_2004}.

The second model (ii) are two interacting high-spins with $S_i=5/2$, which interact via a mix of Heisenberg and Dzyaloshinskii–Moriya (DM) interactions, 
$$J_{ij}^{\alpha \beta} = J \delta_{\alpha \beta} + D \sum_{\gamma} \epsilon_{\alpha \beta \gamma}.$$
The conventional decomposition splits the 25 pairwise interactions into five groups of five, which are applied sequentially. In contrast, there is a dynamical projection scheme for this model with $K=2$ (see SM). This improvement in Floquet engineering efficiency compensates the overhead from introducing a five-qubit gate, leading to slightly higher accessible simulation times. Scaling with $S_i$ is discussed in SM.

The final two models (iii and iv) are composed of spin-2 particles on a square lattice. 
In model (iii), nearest-neighbor spins interact via a Heisenberg and bi-quadratic interaction, $\sum_{k=1}^2 J_k (\mathbf{S}_i \cdot \mathbf{S}_j)^k$.
The conventional Trotter approach requires realizing many four-qubit interactions per spin-2 producing very long Floquet periods. In contrast, dynamical projection can be implemented by applying a single four-qubit gate per edge, resulting in smaller $K$ (see SM).
The multi-qubit implementation is also signficantly more efficient, as the two-qubit decomposition of a four-qubit interaction comes with large overhead, as illustrated in Fig.~\ref{fig:implementation}c.
To highlight the separate contribution from Floquet projection and multi-qubit gates, we compute the effective simulation time for four cases in the table below.

\begin{table}[h]
    \centering
    \begin{tabular}{c|c|c}
    $T_{\mathrm{sc}} \times ||H||_{\mathrm{local}}$ & Two-qubit  gates & Four-qubit gates \\
    \hline
        Trotterization & 0.15 & 1.0\\
        Dynamical Projection & 0.25 & 2.5
    \end{tabular}
    \caption{The quantum simulation's achievable coherence time $T_{\mathrm{sc}}$ in the target Hamiltonian's units $||H||_\mathrm{local}$ (see also~\fref{fig:ham_simulation_comparision}) for four different simulation schemes applied to model (iii). This comparison illustrates the advantages of the techniques we propose here. Hamiltonian engineering based on dynamical projection in combination with efficient, hardware-optimized interactions outperforms alternative approaches based on bare Trotter evolution of two- or multi-qubit gates, as well as dynamical projection just using two-qubit gates.}
    \label{tab:four_square_comparision}
\end{table}

Finally, we consider a model including up to bi-quartic terms $k\leq 4$.
Here, we further consider using an eight-qubit multi-qubit gate to directly realize the interaction between two large-spins (iv). This interaction naturally preserves the symmetry, so dynamical projection is not needed in this case. As such, the speedup comes entirely from the efficiency of the optimized eight-qubit operation, compared against the large overhead associated with a two-qubit decomposition of an eight-qubit interaction.

\subsection{Many-body Spectroscopy}

To complete our simulation framework, we also develop tools for resource-efficient readout of Hamiltonian properties.
First, we illustrate how to compute two-time correlation functions of the form

\begin{equation}
    C_{O,R}(t) = \langle S \vert O(t) R(0) \vert S \rangle,
\end{equation}

\noindent using the circuit in Fig.~\ref{fig:spectroscopy_toy_model}a.
The real and imaginary parts of $C_{O,R}(t)$ are independently accessed by measuring the ancilla in the $X$ and $Y$ basis, respectively~\cite{lin_heisenberg_limited_2022,obrien_quantum_2019,lu_algorithms_2021,wan_randomized_2022}.
More concretely, consider the state of the system right before measurement, including both the ancilla qubit and the system,

\begin{align}
    \ket{\psi_{\mathrm{f}}} = \frac{1}{\sqrt{2}} \left( \ket{0} \otimes U(t)\ket{S} + \ket{1} \otimes U(t) R \ket{S} \right),
\end{align}

\noindent where $U$ is the time-evolution operator.
Measuring  $X \otimes O$ or $Y \otimes O$ results in

\begin{align}
    \braket{X \otimes O }_{\psi_{\rm f}} &= \frac{1}{2}\left(\bra{S} R^{\dagger} O(t) \ket{S} + \bra{S} O(t) R \ket{S} \right), \\
    \braket{Y \otimes O }_{\psi_{\rm f}} &= \frac{i}{2}\left(\bra{S} R^{\dagger} O(t) \ket{S} - \bra{S} O(t) R \ket{S} \right),
\end{align}

\noindent which together gives the full complex-valued $C_{O,R}(t)$ by taking a linear combination of the two,

\begin{align}\label{eq:C_OR}
    C_{O,R} = \langle (X + iY) \otimes O \rangle_{\psi_{\rm f}}.
\end{align}

\noindent For observables $O$ diagonal in the measurement basis, $C_{O,R}$ can be efficiently estimated in parallel from snapshots.
During the $i$-th run, let $\mu^{(i)} = \{ x,y \}$ be the randomly sampled ancilla measurement basis, and $a^{(i)} = \{0,1\}$, and $\vert b^{(i)} \rangle$ be the ancilla and system measurement outcomes respectively.
Then, the estimator can be written as

\begin{align}{\label{eqn:COR_estimator}}
    \overline{C_{O,R}(t)} = \frac{1}{M}\sum_{i=1}^{M} 2\sigma(\mu^{(i)}, a^{(i)}) \langle b^{(i)} \vert O \vert b^{(i)} \rangle
\end{align}

\noindent where $\sigma$ is a function taking on the values

\begin{align}
    &\sigma(x,0) = +1, &\sigma(x,1) = -1, \nonumber\\
    &\sigma(y,0) = +i, &\sigma(y,1) = -i.
\end{align}

\noindent and $\vert b^{(i)} \rangle$ is the measured projected state.

These measurements can be used to compute the operator-resolved density of states \eqref{eqn:raw_DOS}, which can be rewritten as follows

\begin{align}
    D^{A}(\omega) &= \int dt\, e^{i \omega t} \sum_n \tr[A \vert n \rangle e^{-i \epsilon_n t} \langle n \vert] \nonumber \\
    &= \int dt\, e^{i \omega t} \tr[A U(t)],
\end{align}

\noindent where we have replaced $\delta(\omega - \epsilon_n) \rightarrow \int dt \, e^{i(\omega - \epsilon_n)t}$.
In practice, we will sample evolution times $t$ from a probability distribution $p(t)$, such that the integral is normalized to one when $\omega = \epsilon_n$, \ie $\int dt = \int_{-\infty}^{\infty} p(t) dt = 1$.
To arrive at \eqref{eq:dos_detailed}, we can replace the trace with an average over probe states~\cite{zintchenko_randomized_2016}

\begin{align}{\label{eqn:D_RA_raw}}
    \tr[A U(t)] = \tilde{\mathds{E}}_{R \sim \mathcal{R}} \langle R \vert A U(t) \vert R \rangle,
\end{align}

\noindent where $\tilde{\mathbb{E}}_{R \sim \mathcal{R}} = \mathrm{Tr}[{\mathds{1}}] \mathbb{E}_{R \sim \mathcal{R}}$ is a normalized expectation value, and $\mathrm{Tr}[\mathds{1}]$ is the dimensionality of the Hilbert space.
This is valid, as long as the ensemble forms a 2-design, 

\begin{align}
    \mathbb{E}_{R \sim \mathcal{R}} \vert R \rangle \langle R \vert = \frac{\mathds{1}}{\tr[{\mathds{1}}]},
\end{align}

Observables such as \eqref{eqn:D_RA_raw} can in principle be computed via a modified Hadamard test by applying controlled-time evolution (see Ref.~\cite{lu_algorithms_2021}).
Since time-evolution is generally the most costly step, we avoid the overhead associated with controlled-evolution and instead utilize a reference state $\vert S \rangle$ with simple time-evolution.
In particular, we select an ensemble of observables $O_s$ such that 

\begin{align}
    \frac{1}{\mathcal{N}(O_s)}\sum_s O_s U(t) \vert S \rangle \langle S \vert U^{\dagger}(t) O_s = \mathds{1}.
\end{align}

\noindent The normalization factor $\mathcal{N}(O_s)$ depends on the choice of ensemble.
Then, we can insert this resolution of the identity into \eqref{eqn:D_RA_raw} to get \eqref{eq:dos_detailed}.
In Fig.~\ref{fig:spectroscopy_toy_model} and Fig.~\ref{fig:OEC}, we consider the polarized reference state $\vert S \rangle = \vert {0} \rangle^{\otimes N}$ which is an exact eigenstate of the Heisenberg Hamiltonian, and the ensemble of Pauli-$X$ operators $O_s = X_s = \bigotimes_{i=1}^N (X_i)^{s_i}$ where $s$ is an $N$-bit string.
This satisfies the condition, and has $\mathcal{N}(X_s) = 1$ since $X_s \vert 0 \rangle^{\otimes N}$ is an orthonormal basis.
For generic reference states $\vert S \rangle$ prepared by applying $S$ to $\vert 0 \rangle^{\otimes N}$, the ensemble $O_s = S X_s S^{\dagger}$ satisfies the condition, and can be measured by applying the inverse preparation circuit $S^{\dagger}$ before measuring in the $X$-basis.
Lastly, an ensemble which is independent of the reference state, is the set of Pauli strings $P_s = \bigotimes_{i=1}^{N} \sigma_i^{s_i}$, where $s$ is a base-four string, and $\sigma^{s_i}$ denotes the four Pauli operators $I,X,Y,Z$; this can be accessed with randomized measurements (see SM), and has a normalization factor $\mathcal{N}(P_s) = 2^{N}$.

\subsubsection{Efficient estimation of $D^{A}_R(t)$ from snapshots}
One of the key advantages of our approach, is the ability to compute many estimators $D^{A}_R(t)$ from one dataset, including complex operators such as projectors onto spin-sectors $A = P_S$.
This enhances the sample efficiency and reduces simulation time requirements, which are crucial limited resources in quantum computation.

To estimate density-of-states in a sample efficient way, we introduce a classical co-processing algorithm which utilizes knowledge of the prepared reference state.
In particular, we discuss the estimation procedure for

\begin{align}\label{eqn:DAR_exact}
    D^{A}_R(t) &= \sum_s \underbrace{\langle R \vert A O_s U(t) \vert S \rangle}_{\rm{classical}} \underbrace{\langle S \vert U^{\dagger}(t) O_s U(t) \vert R \rangle}_{\rm{quantum}} \nonumber\\
    &= e^{-i E_S t} \sum_s \langle R \vert A O_s \vert S \rangle \langle S \vert O_s(t) \vert R \rangle
\end{align}

\noindent which becomes \eqref{eq:dos_detailed} after averaging over times $t$ and perturbations $R$.
In the second line, we assumed $\vert S \rangle$ is a known eigenstate, so $U(t) \vert S \rangle = e^{-i E_S t} \vert S \rangle$, and the classical part becomes equivalent to a zero-time correlation function.
One challenge is that the sum in~\eqref{eqn:DAR_exact} involves exponentially many observables.
However, by using the estimator \eqref{eqn:COR_estimator} this reduces to a sum over $M$ snapshots.
In particular, for the polarized reference state, and Pauli-$X$ measurements, a simple calculation (see SM) shows that

\begin{align}\label{eqn:DRA_estimator}
    \overline{D^{A}_R(t)} = \frac{1}{M} \sum_{i=1}^{M} e^{-i E_S t} 2^{N/2} \sigma(\mu^{(i)}, a^{(i)}) \langle R \vert A \vert b^{(i)} \rangle 
\end{align}

\noindent is an unbiased estimator. 
The variance of this estimator is independent of system size for appropriately chosen ensembles $\mathcal{R}$, and scales as $O(1/\sqrt{M})$.
While $C_{O_s,AR}^*(0)$ could instead be estimated by measuring the unevolved state, this would produce an estimator that converges very slowly, leading to a sample complexity that is exponential in $N$.
However, classical methods can compute $C_{O_s,AR}^*(0)$ with no error, making the procedure much more sample-efficient.
In the SM, we discuss how such calculations can be efficiently performed for any pair of reference and probe states with an efficient matrix product state (MPS) description, and any observable $A$ with a matrix product operator (MPO) description. 
Further, we show that projectors $P_S$ onto spin-sectors can be written as MPOS with $\mathrm{poly}(N)$ bond-dimension, making evaluation of $D^{P_S}(\omega)$ classically efficient and scalable.
We further discuss how entangled measurements between two systems could be used to estimate these correlators efficiently, even when $\vert S \rangle$ does not have a known classical description.

\subsubsection{Thermal expectation values}

The operator-resolved density of states can be used to compute thermal expectation values via~\cite{lu_algorithms_2021}

\begin{align}\label{eq:finite_temp}
    \langle A \rangle_{\beta} = \frac{\int e^{-\beta \omega} D^{A}(\omega)d\omega }{\int e^{-\beta \omega} D^{\mathds{1}}(\omega)d\omega}.
\end{align}

\noindent For example, to compute the magnetic susceptibility, we simply select the operator $A = \beta (S^z)^2$, where $\beta = 1/T$ is the inverse temperature.
Interestingly, this method of estimating thermal expectation values is insensitive to uniform spectral broadening of each peak, due to a cancellation between the numerator and denominator (see SM).
However, it is highly sensitive to noise at low $\omega$, which is exponentially amplified by $e^{- \beta \omega}$. 
To address this, we estimate the SNR for each $D^{A}(\omega)$ independently, and zero-out all points with SNR below three times the average SNR. This potentially introduces some bias by eliminating peaks with low-signal, but ensures the effects of shot-noise are well controlled.

\subsection{Noise Modelling}

To quantify the effect of noise on the engineered time-dynamics, we simulate a microscopic error model by applying a local depolarizing channel with an error probability $p$ at each gate.
This results in a decay of the obtained signals for the correlator $D^{A}_R(t)$. 
The rate of the exponential decay grows roughly linearly with the weight of the measured operators (see Extended Data Fig.~\ref{fig:supp_noise_fig}).
This scaling with operator weight can be captured by instead applying a single depolarizing channel at the end of the time-evolution, with a per-site error probability of $\gamma t$ with an effective noise rate $\gamma$.
This effective $\gamma$ also scales roughly linear as a function of the single-qubit error rate per gate $p$ (see Extended Data Fig.~\ref{fig:supp_noise_fig}).

\subsection{Scaling the approach}

Quantum simulations are constrained by the required number of samples and the simulation time needed to reach a certain target accuracy. These factors are crucial for determining the size of Hamiltonians which can be accessed for particular quantum hardware.

Focusing on a single gapped eigenstate we determine the number of snapshots $C_M$ needed to distinguish a spectral peak from noise (Fig.~\ref{fig:samp_complexity_scaling}).
The signal arises from the overlap of the probe states with the target eigenstate. The noise is given by the variance of the estimator~\eqref{eqn:DRA_estimator}, and decays as $\epsilon \sim M^{-1/2}$. 
For certain ensembles of probe states, the variance can be made system size independent (see SM).
However, a random probe state will have exponentially vanishing overlap with any specific eigenstate.
One approach to mitigate this, is to initialize probe states with higher overlap.
In Fig.~\ref{fig:state_ovlps_comp}b we show that for a spin-1 AFM chain a simple bond-dimension two MPS can outperform product states by orders-of-magnitude in ground-state estimation. 
While bond-dimension two states can be efficiently prepared with simple circuits of two qubit gates, more general ansatze can also be efficiently realized using the simulation techniques described here~\cite{haghshenas_variational_2022,foss-feig_holographic_2021}. Optimized ansatze could be further combined with importance sampling~\cite{mcclean_theory_2016}, to improve the sample-efficiency of computing finite-temperature or excited state properties (see SM).

The simulation time $T_{\mathrm{max}}$ will depend on the required spectral resolution, which does not scale with system size for a gapped eigenstate.
However, the rate of spectral broadening depends sensitively on the weight of measured observables (Fig.~\ref{fig:supp_noise_fig}).
When the reference state is high in energy, such as the polarized state for an AFM chain, the relevant observables typically have extensive weight, requiring $T_{\mathrm{max}} \sim N$ to maintain constant spectral resolution.
In contrast, preparing a low-energy reference state, such as the ground-state $\vert S \rangle = \vert GS \rangle$, allows coupling to other low-energy states using low-weight operators. This results in a noise-resilient and system-size independent procedure (Fig.~\ref{fig:state_ovlps_comp}).
We further note that ground-state preparation can be approximate, which would result in additional spectral broadening in the computation of $D^{A}(\omega)$.
While the spectral resolution requirements should also grow as the gap shrinks, we have illustrated that operator-resolution can mitigate this in certain settings (e.g. Fig.~\ref{fig:OEC} and Fig.~\ref{fig:extended_systems}). As such, understanding the general capabilities of this approach is an interesting direction for continued research.

\subsection{OEC Hamiltonians}

The two candidates for the closed S2 state of the oxygen evolving complex (OEC) are parameterized with Heisenberg models $H=-\sum_{ij} J_{ij} \hat{\mathbf{S}}_i\cdot\hat{\mathbf{S}}_j$~\cite{krewald_magnetic_2013, krewald_spin_2016}. The coupling constants for the two cases used to generate the data presented in~\fref{fig:OEC} are summarized in Table~\ref{tab:couplings}.
The corresponding spin sizes of the different sites are $S_1=3/2$, $S_2=3/2$, $S_3 = 3/2$, $S_4=2$.
\begin{table}[h]
    \centering
    \begin{tabular}{C{1.5cm} C{1cm} C{1cm} C{1cm} C{1cm} C{1cm} C{1cm}}
        \hline
         & $J_{12}$ & $J_{13}$ & $J_{14}$ & $J_{23}$ & $J_{24}$ & $J_{34}$ \\
        \hline
         $\mathrm{\mathbf{S_2H}}$\textbf{-1b} & 30.5 & 12.9 & 4.5 & 36.5 & 1.3 & -7.3 \\
        $\mathrm{\mathbf{S_2H}}$\textbf{-2b} & 32.6 & 11.7 & 4.0 & 37.3 & 1.5 & -2.6\\
        \hline
    \end{tabular}
    \caption{Exchange coupling constants in units of wavenumbers $(\mathrm{cm})^{-1}$ for the two different candidates for the $S_2$ stage of the OEC used in~\fref{fig:OEC}.}
    \label{tab:couplings}
\end{table}

Additional information about eigenstates can be calculated by choosing the operator $A$ in the operator resolved density of states appropriately, and multiplying by a narrow ``band-pass'' filter in Fourier space to isolate a small set of frequencies~\cite{lu_algorithms_2021,schuckert_probing_2022}.
For example, we investigate the total spin of the cubane sub-unit, \ie the three magnetic sites supported on opposite vertices of the cube, using $A = (\hat{\bf{S}}_1 + \hat{\bf{S}}_2 + \hat{\bf{S}}_3)^2$, and compute

\begin{align}
    A(\omega) = \frac{D^{A}(\omega)}{D^{\mathds{1}}(\omega)}
\end{align}

\noindent evaluated at energies $\omega_n$ of the eigenstates, which can be extract from peaks in the spectral functions (Extended data fig.~\ref{fig:supp_OEC_extra}).
We further use spin-projection to improve the estimate in the presence of broadening. For example, if the peak at $\omega_n$ occurs in spin-sector $P_S$, we insert the projector $P_S$ in the numerator and denominator, $A(\omega_n) = \frac{D^{A P_S}(\omega_n)}{D^{P_S}(\omega_n)}$.

\subsection{2D Heisenberg calculations}

The square lattice Heisenberg calculation was performed on a large ($L \times L$) system, with Hamiltonian

\begin{align}
    H_{2D} = -J \sum_{\mathbf{r}} \left( \hat{\mathbf{S}}_{\mathbf{r}} \cdot \hat{\mathbf{S}}_{\mathbf{r} + \hat{x}} + \hat{\mathbf{S}}_{\mathbf{r}} \cdot \hat{\mathbf{S}}_{\mathbf{r} + \hat{y}} \right)
\end{align}

\noindent We measure the Green's function $G(\mathbf{r},t) = \langle S \vert X_{\mathbf{r}}(t) X_{\mathbf{0}} \vert S \rangle$ from a polarized reference $\vert S \rangle^{\otimes L^2}$~\cite{knap_probing_2013}.
Since $S^z$ is conserved under the dynamics $G(\mathbf{r},t)$ is classically simulated by restricting it to the space containing the $\vert S \rangle$ and single spin-flip states $X_{\mathbf{r}} \vert S \rangle$, which has dimension $L^2+1$.
We evolve under equally spaced times up to $J T_{\rm{max}} = 8$, and select $L=23$ which is large enough such that $G(r,t)$ vanishes far from the boundaries.
Therefore, by letting $G(r,t)=0$ outside the simulated region, this provides a good approximation for the $L \rightarrow \infty$ limit.
As such, we define projectors onto (unnormalized) plane-wave states $P_\mathbf{k} = \vert \mathbf{k} \rangle \langle \mathbf{k} \vert$, where $\langle \mathbf{k} \vert X_{\mathbf{r}}\vert S \rangle = e^{-i \mathbf{k} \cdot \mathbf{r}}$.
Then, $D^{P_\mathbf{k}}(\omega)$ can be written as

\begin{align}
    D^{P_\mathbf{k}}(\omega) &= \int dt e^{i (\omega - \omega_0) t}\sum_\mathbf{r} \underbrace{\langle S \vert X_{\mathbf{0}} P_k  X_\mathbf{r} \vert S \rangle}_{e^{i \mathbf{k} \cdot (\mathbf{r} - \mathbf{0})}} \underbrace{\langle S \vert X_\mathbf{r}(t) X_\mathbf{0} \vert S \rangle}_{G(r,t)}
\end{align}

\noindent which reduces to the Fourier transform $G(k,\omega)$ when the energy of the polarized state $\omega_0$ is set to zero.
Plotting this for a continuous set of $k$ and $\omega$ produces the spectral weight depicted in Fig.~\ref{fig:extended_systems}.
This further shows that finite-size systems are sufficient to simulate extended systems at finite evolution times.

By computing the peak value of $\omega$ for each $k$ in the vicinity of the single-particle excitations, we estimate the dispersion relation $\omega(k)$ associated with a single spin-flip excitation. Computations of the many-particle Green's function could be performed similarly by applying multi-site perturbations, to extract finite-temperature properties and characterize the interactions between the quasi-particles.


\clearpage
\onecolumngrid
\renewcommand\figurename{Extended Data -- FIG.}

\begin{figure}
    \centering
    \includegraphics{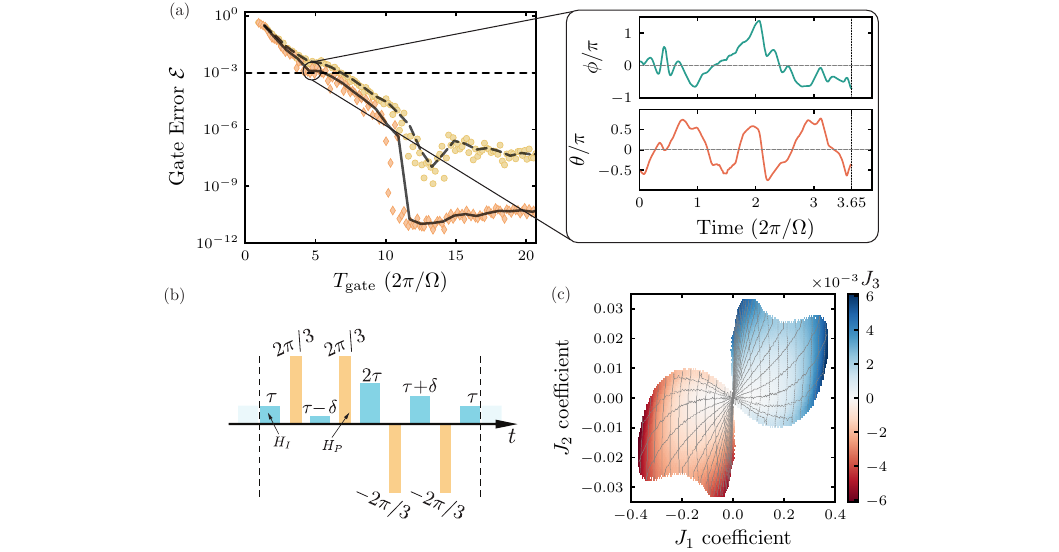}
    \caption{\textbf{Gate optimization procedure and variational Hamiltonian Engineering}. (a) To find smooth gates we perform the GrAPE optimization procedure in two steps. In the first step, we penalize rapid changes in the pulse profile by introducing an extra term in the cost-function. In this case, the resulting relationship between ideal gate-error vs. time (yellow circles) saturates around $10^{-6}$. For the second step we initialize the search with the smooth gates found in the previous step, which are re-optimized by removing the smooth penalty. This significantly reduces the ideal gate error (orange). The data are shown for a $N=4$ qubit cluster. 
    Since the first step already confined the problem into a subspace of the search space with smooth gates, the resultant pulses also remain smooth after the second step. On the right we show an example smoothened pulse profile for the hyperfine angle $\theta$ and the Rydberg phase $\phi$ (see Methods) with an ideal gate error rate $\mathcal{E} = 10^{-3}$. 
    (b) Higher order interaction terms can be controllably engineered via a simple modification of the $K=6$ Floquet projection sequence. For example, reducing the second time-step and increasing the fourth time-step by the same amount, $\delta$ preserves the target Hamiltonian at leading order. (c) By tuning $\tau$ and $\delta$, for a system of two interacting spin-3/2's, a very large family of coefficients $J_1, J_2, J_3$ in the general Hamiltonian $H_T = J_1 (\mathbf{S}_1 \cdot \mathbf{S}_2) + J_2 (\mathbf{S}_1 \cdot \mathbf{S}_2)^2 + J_3 (\mathbf{S}_1 \cdot \mathbf{S}_2)^3$ can be engineered. In particular, the roughly horizontal gray lines correspond to constant $\delta$ grid lines, and roughly vertical gray lines correspond to values with constant $\tau$. We see that changing $\delta$ primarily modifies $J_2$ and $J_3$, while changing $\tau$ primarily modifies $J_1$, consistent with our analysis that $\delta$ picks out certain higher-order terms. We further note that the especially interesting AKLT family, with $J_2/J_1 = 116/243$ and $J_3/J_1 = 16/243$ lies among the family of efficiently engineerable interactions.}
    \label{fig:supp_gates_and_higher_order}
\end{figure}

\begin{figure}
    \centering
    \includegraphics{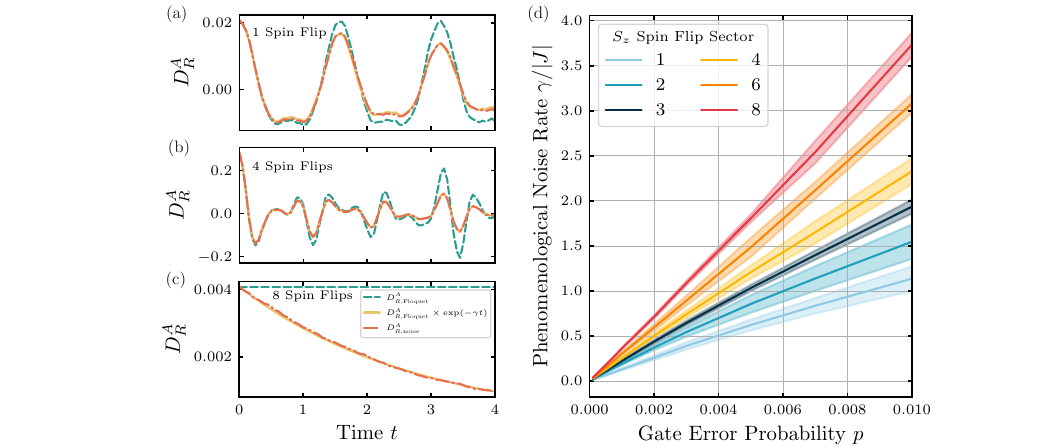}
    \caption{\textbf{Gate-level noise simulations}. (a)-(c) Time evolution of the correlations $D^{A}_R(t)$ for a polarized reference state $\vert S \rangle = \ket{0}$ and $A$ chosen to be projectors onto different $S^z$ sectors. This involves measuring operators of fixed weight. We see that the simulation of gate-level noise modelled as a depolarizing channel with a gate error probability $p=0.001$ (orange dash dotted curve) matches well with the dynamics obtained by adding an additional phenomenological noise $\propto \exp(-\gamma t)$ with rate $\gamma = (0.128, 0.251, 0.36)$ for $N_\mathrm{spin\,flip} = (1,4,8)$ (solid yellow lines) to the loss-less time evolution $D^A_{R, \mathrm{Floquet}}$ (green  dashed). (d) Relationship between the optimal phenomenological noise rate $\gamma$ and the gate error probability $p$ for different spin flip sectors. We see a roughly linear-relationship of the decay rate $\gamma$ with the gate error rate $p$ and operator weight $S_z$. All data are obtained using time-steps $J\tau = 0.05$.
    }
    \label{fig:supp_noise_fig}
\end{figure}

\begin{figure*}
    \centering
    \includegraphics{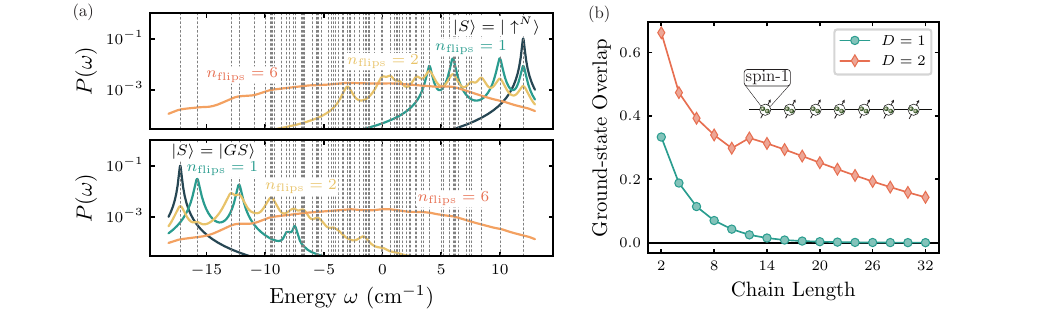}
    \caption{\textbf{Noise susceptibility and eigenstate overlaps}. (a) Density-of-states for a spin-1 AFM chain, computed from a polarized reference state $\vert S \rangle = \vert {0} \rangle^{\otimes N}$ (top), and ground state reference $\vert S \rangle = \vert GS \rangle$ (bottom). The spectrum is separated into sectors distinguished by their operator weight from $\vert S \rangle$. For the polarized state, these correspond to sectors with well-defined $S^z$. For the ground-state, each sector is orthogonalized with respect to lower-weight sectors.
    Each sector is phenomenologically broadened by $e^{-\gamma t n_{\mathrm{flips}}}$ to simulate the operator-weight dependence of decoherence. 
    When computing low-energy properties, the ground-state reference is more robust to noise, since the low-energy eigenstates can be reached with lower-weight operators. (b) The amplitude of the corresponding spectral peak is determined by the eigenstate overlap. We analyze the ground-state overlap for an AFM spin-1 chain performing DMRG for low bond dimensions ($D=1,2$) and find that it is significantly larger for bond-dimension $D=2$ matrix product states (red diamonds) compared to bond-dimension $D=1$,~\ie mean-field states (green circles). Interestingly, the ground state overlap decays slower with the chain length for $D=2$ indicating that the fidelity density is large. This feature makes low bond dimension states a promising direction for efficient state preparation within our scheme since they can efficiently be decomposed into short circuits.
    }
    \label{fig:state_ovlps_comp}
\end{figure*}

\begin{figure*}
    \centering
    \includegraphics{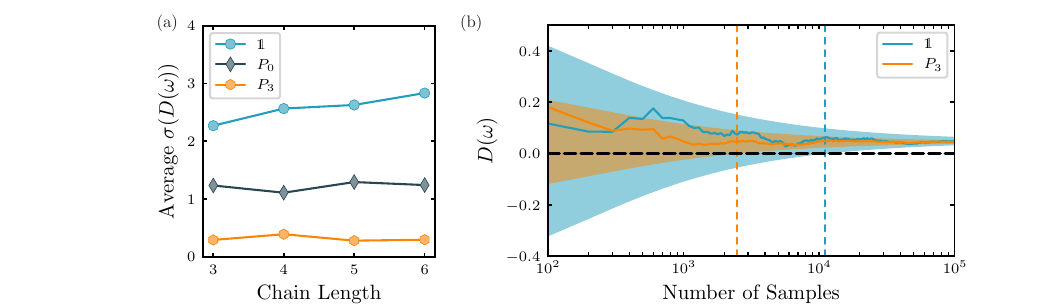}
    \caption{\textbf{Shot noise scaling with system size and convergence with number of snapshots}. (a) Numerically computed standard deviations of the estimator \eqref{eq:dos_detailed} for the density of states $D^{A}(\omega)$ of a spin-1 AFM chain, for different chain lengths and observables. Here, we consider a polarized reference state $\vert S \rangle$ and random single-qubit rotations $R$ for the controlled-perturbation.
    For the bare density of states the standard deviation slowly scales with the chain length (blue circles). For the projectors into the zero (gray diamonds) and three (orange circles) spin-flip sectors the standard deviation is consistent with being independent of system size. (b) Convergence of estimator with number of samples for two different observables. Shaded regions correspond to $2\sigma$ error bars around the mean, and decrease with the number of measurements as $1 / \sqrt{M}$. Dark lines are running averages for a specific sampled dataset.
    The sample complexity is defined as the number of samples needed such that the error bars around the mean do not include zero. After this point, spectral peaks can be reliably distinguished from noise.}
    \label{fig:samp_complexity_scaling}
\end{figure*}

\begin{figure}
    \centering
    \includegraphics{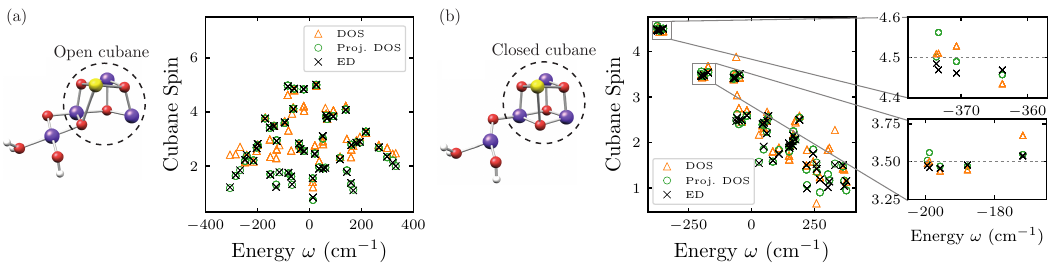}
    \caption{\textbf{Additional observables calculated for OEC}. By choosing $A$ to be the local cubane spin $S_{123}^2$, and calculating $D^{A}(\omega)$ at each of the peaks in the density of states, we can infer the local structure of spin correlations within each eigenstate. (a) The cubane spin for each eigenstate for the open cubane configuration ($\mathrm{S_2}$H-1a). This model has coefficients: $J_{12} = -15.8, J_{13} = 1.9, J_{14}, J_{23} = 23.1, J_{24} = 1.9, J_{34} = -13.9$ in $\mathrm{cm}^{-1}$, and spin-sizes $S_1=2, S_2=S_3=S_4=3/2$. The triangles, circles and crosses are obtained using three different methods. The crosses indicate the exact value obtained via exact diagonalization (ED), the green circles indicate the values obtained via the spin resolved (projected) density of states (DOS), and the orange triangles are obtained by the bare DOS. The projected DOS, which is able to resolve individual eigenstates, matches well with the ED result. In contrast the bare DOS, which has overlapping peaks (see Fig.~\ref{fig:OEC}), does not capture the expectation values accurately. (b) Same as in (a) but for the closed cubane configuration $\mathrm{S_2}$H-1b. This model has $S_1=S_2=S_3=3/2$ and $S_4=2$. We see that for the lowest-lying cluster, all spins in the cubane subunit are approximately polarized with a total spin value $S = 9/2 = 4.5$. In contrast, the second cluster of eigenstates seems to differ by a single spin flip as indicated by the value $S = 7/2 = 3.5$.}
    \label{fig:supp_OEC_extra}
\end{figure}

\pagebreak

\clearpage
\onecolumngrid
\begin{center}

\newcommand{\beginsupplement}{%
        \setcounter{table}{0}
        \renewcommand{\thetable}{S\arabic{table}}%
        \setcounter{figure}{0}
        \renewcommand{\thefigure}{S\arabic{figure}}%
     }
\textbf{\large Supplemental Material}
\end{center}
\newcommand{\beginsupplement}{%
        \setcounter{table}{0}
        \renewcommand{\thetable}{S\arabic{table}}%
        \setcounter{figure}{0}
        \renewcommand{\thefigure}{S\arabic{figure}}%
     }
\setcounter{equation}{0}
\setcounter{figure}{0}
\setcounter{table}{0}   
\setcounter{page}{1}
\makeatletter
\renewcommand{\theequation}{S\arabic{equation}}
\renewcommand{\thefigure}{S\arabic{figure}}
\renewcommand{\bibnumfmt}[1]{[S#1]}
\renewcommand{\citenumfont}[1]{S#1}
\vspace{0.8 in}
\newcommand{\D}{\Delta}
\newcommand{\tD}{\tilde{\Delta}}
\newcommand{\K}{K_{PP}}
\newcommand{\bn}{\bar{n}_P}
\newcommand{\G}{\Gamma}
\newcommand{\LH}{\underset{L}{H}}
\newcommand{\HL}{\underset{H}{L}}
\vspace{-1in}

\section{Hamiltonian Engineering}

\subsection{Variational spin-gate optimization.}

An exciting route to efficiently generating more complex interactions is to go beyond average Hamiltonian engineering.
In particular, we introduce a $p$-dependent evolution time $\tau_p$, and consider tuning the parameters $\tau_p$ and $\theta_p$ in the pulse sequence \eqref{eqn:floquet_dyn_proj} to control certain higher-order terms, with no additional gate overhead.
As a concrete example (see also Extended data Fig.~\ref{fig:supp_gates_and_higher_order}), we consider two spin-3/2 particles, evolved by $H_I = \frac{9J}{4} \hat{\bf{s}}_{1,1} \cdot \hat{\bf{s}}_{2,1}$ and a $K=6$ Floquet sequence with parameters
\begin{align}
    \Theta &= (0, 2\pi/3, 4\pi/3, 4\pi/3, 2\pi/3, 0) \\
\boldsymbol{\tau} &= (\tau, \tau-\delta, \tau, \tau, \tau+\delta, \tau).
\end{align}
We find that this sequence naturally engineers an SU(2) symmetric Hamiltonian with Heisenberg interactions, but also bi-quadratic and bi-cubic terms
\begin{align}\label{eqn:H2_extended}
    H_{2,extended} = J_1 (\hat{\bf{S}}_1 \cdot \hat{\bf{S}}_2) + J_2 (\hat{\bf{S}}_1 \cdot \hat{\bf{S}}_2)^2 + J_3 (\hat{\bf{S}}_1 \cdot \hat{\bf{S}}_2)^3.
\end{align}

To characterize the values of $J_1,J_2$, and $J_3$ which can be accurately realized, we simulate the action of $U_F$ on the state $\rho_\mathrm{sym}$ which is maximally mixed in the encoded subspace, for various pairs of values $\tau, \delta$.
Then, as long as long as the average population leaving the subspace is below $10^{-4}$, we project $U_F$ into the encoded subspace, compute the effective Hamiltonian $H_F = -i \log(U_F)$ by exact diagonalization, and fit the coefficients $J_1, J_2, J_3$ by solving a set of linear equations.
The resulting family of engineerably coefficients forms a 2D surface in a 3D space, and is visualized in Extended data Fig.~\ref{fig:supp_gates_and_higher_order}b.
Interestingly, the AKLT family of Hamiltonians~\cite{affleck_rigorous_1987}
\begin{equation}
H_{\mathrm{AKLT}, 3/2} = \sum_{i} \sum_{k=1}^{3} J_k^\mathrm{AKLT} \left( \vec{S}_i \cdot \vec{S}_{i+1} \right)^k,
\end{equation}
lies within the space of engineerable Hamiltonians.
Here, the coefficients $J_k^\mathrm{AKLT}$ are chosen such that $H_{\mathrm{AKLT}, 3/2}$ is equivalent to the projection Hamiltonian $H_P$ on six qubits. 
In other words the groundstate of $H_{\mathrm{AKLT}, 3/2}$ is the total spin-3 subspace, and is separated by a constant gap to the other lower spin subspaces.

The procedure outlined above could also be applied to anisotropic Hamiltonians $H_I$, to try and engineer the full-space of two-spin operations.
Extensions beyond pairwise interactions could also be considered, for example to efficiently realize three-spin interactions as in Ref.~\cite{kalinowski_non-abelian_2023}.
We note that, higher-order interactions in our approach emerge similarly to those in realistic materials.
In our case they arise from virtual total-spin fluctuations, whereas in a material they arise from virtual charge-fluctuations.

\subsection{Procedures for Multi-Qubit Gate Optimization}

The Hamiltonian engineering approach described in this work stands out for its hardware efficiency, owing to the strategic use of efficient multi-qubit gates. 
To perform optimization efficiently, we first introduce a low-dimensional basis for each cluster, leveraging the permutation symmetry of the blockade interaction.
Then, we illustrate two complementary approaches to determing pulse profiles which realize the target gate.
First, we show how to use the GrAPE algorithm to find smooth off-diagonal gates where the pulses $\Omega_q(t)$ and $\Omega_r(t)$ are applied simultaneously.
Second we describe the two-stage optimization used to identify alternating sequences of diagonal multi-qubit interaction gates and single-qubit gates to generate the target off-diagonal operations.

\subsubsection{Low-dimensional construction of blockade cluster basis}

To efficiently perform numerical optimization of the pulse sequences $\Omega_q(t)$ and $\Omega_r(t)$ (see below), we use a low-dimensional representation of the Hilbert space, leveraging the permutation-symmetry of the time-dependent Rydberg Hamiltonian~\eqref{eqn:Ryd_Ham}. In the blockade approximation, states with more than one Rydberg excitation within the cluster are disallowed.
When the Rydberg Hamiltonian is projected into this manifold, it becomes an all-to-all interacting ``PXP'' model, where a spin-flip is only allowed if all other atoms are in the qubit ($\vert {0} \rangle, \vert {1} \rangle$) states.
Define $P_i$\,$=$\,$\vert {1} \rangle_i \langle {1} \vert + \vert {0} \rangle_i \langle {0} \vert$ to be the projector onto the qubit state at site $i$, and $P$\,$=$\,$\prod_{i=1}^n$\,$P_i$ to be the global projector.
Then, the constrained Hamiltonian can be written as
\begin{align}
    H_{\mathrm{proj}} = \frac{\Omega_{q}(t)}{2} \Sigma_{q}^+ + \frac{\Omega_{r}(t)}{2} \Sigma_{r}^+ P + h.c.
\end{align}
where we have defined collective raising and lowering operators $\Sigma_{q}^+$\,$=$\,$\sum_i \vert {0} \rangle_i \langle {1} \vert$ and $\Sigma_{r}^+$\,$=$\,$\sum_i \vert r \rangle_i \langle {0} \vert$.
This is a good approximation for the dynamics at leading order in $\Omega_r/V$, and has an emergent permutation symmetry. In particular, notice that $H_{\mathrm{proj}}$ is invariant under exchange of any two sites.
In general, the Hilbert space of an $N$-atom cluster has $3^N$ states, and the blockade projection further reduces this to an $O(2^N)$ scaling.
However, to efficiently perform numerical optimization for larger clusters, we construct a basis whose dimension grows quadratically in $N$ by leveraging the permutation symmetry of $H_{\mathrm{proj}}$.

We start by defining a class of symmetric states which are fully specificed their occupation numbers $\vert n_{{1}}, n_{{0}}, n_r\rangle$.
Each state is  constructed by taking a symmetric superposition of all configurations with the corresponding occupations. 
For example, the $N=2$ triplet state is given by 
$$\vert 1,1,0\rangle = (\vert {0}{1}\rangle + \vert {1}{0}\rangle)/\sqrt{2}.$$
More generally, the states spanned by $\vert n_{{1}}, n_{{0}},0 \rangle$ form the degenerate ground space of $\hat{\mathbf{S}}^2$, and have total spin $N/2$ (where $N = n_{{1}} + n_{{0}}$).
Excited states of $S^2$ live outside the symmetric space, and have total spin ${<}N/2$; thus,
to form the lowest-lying excited states, i.e., states with total spin $N/2$\,$-$\,$1$, we need to expand our basis.

In general, there are $N$\,$-$\,$1$ independent irreducible representations (irrep) with $S=N/2-1$~\cite{arecchi_atomic_1972}.
One irrep can be formed by constructing states that are symmetric under permutation of spins $i=1,...,N-1$, but orthogonal to the fully symmetric state.
A simple way to construct a basis spanning this $S=N/2-1$ irrep and the symmetric $S=N/2$ irrep is to take a tensor product of symmetric spaces on $N-1$ atoms and $1$ atom. 
To do this, we split the atoms into two groups, $A$ and $B$, and label states by six numbers $\vert n_{{1}}^{A},n_{{0}}^{A},n_r^{A},n_{{1}}^{B},n_{{0}}^{B},n_r^{B}\rangle$, associated with the occupation of each group.
Therefore, states must satisfy $n_{{1}}^{A} + n_{{0}}^{A} + n_r^{A} = N-1$ and $n_{{1}}^{B} + n_{{0}}^{B} + n_r^{B} = 1$.
The blockade constraint is imposed by further requiring $n_r^{A} + n_r^{B} \leq 1$.
Then, matrix elements of $H_{\mathrm{proj}}$ can be efficiently constructed in this manifold. More generally, we can construct a basis which spans irreps of spin $S=N/2-k$,$S=N/2-k+1$, to $S=N/2$, by splitting the system into $N-k$ atoms and $k$ atoms. Simulations are still efficient, since the dimensionality scales as $\Theta(N^2)$ when $k=\lfloor N/2 \rfloor$.

\subsubsection{Simultaneous drive -- Smooth gates with GrAPE}

The primary tool we utilize to identify fast, hardware-optimized gate operations is gradient ascent pulse engineering (GrAPE)~\cite{khaneja_optimal_2005}, which has recently been utilized to develop high-fidelity diagonal gates based on Rydberg driving~\cite{jandura_time-optimal_2022,evered_high-fidelity_2023}.
Here, we extend the optimization results to include off-diagonal gates based on simultaneous Rydberg $\Omega_r(t)$ and hyperfine $\Omega_q(t)$ driving.
The core of the algorithm involves approximating time-dependent evolution under $H(t)$ with discrete, Trotterized time evolution.
The time-dependent profile we consider has two parameters, a phase on the Rydberg drive $\Omega_{r}(t) = \Omega_r e^{-i \phi(t)}$, and a hyperfine rotation angle $\theta(t)$ defined by $\Omega_{q}(t) = \partial_t \theta(t)$.
The action of the continuous pulse is approximated by discretizing evolution of $n_T$ steps with step-size $\tau$ each, resulting in a unitary of the form
\begin{align}
    U_{F}(\theta,\phi)=\prod_{k=1}^{n_{T}}e^{i\theta(k\tau)\left(\Sigma_{q}^{+}+\Sigma_{q}^{-}\right)}e^{-i\tau\left(e^{i\phi(k\tau)}\Sigma_{r}^{+}+h.c.\right)}e^{-i\theta(k\tau)\left(\Sigma_{q}^{+}+\Sigma_{q}^{-}\right)}.
\end{align}
Optimization is performed by fixing $T = n_T \tau$, and tuning discrete set of variational parameters $\theta(k\tau),\phi(k\tau)$ by computing the distance between $U_F$ and the target unitary $U_{\mathrm{target}}$.
Since the initial and final states should be fully supported in the qubit manifold, we evaluate the distance between $U_{\mathrm{target}}$ and $U_F$ within this manifold.
Then, we can measure the gate fidelity using the expression
\begin{align}
    \mathcal{F}(\{ \theta, \phi \}) = \frac{N_{\mathrm{q}} + |\sum_q \langle q \vert U_{\mathrm{target}}^{\dagger} U_F(\theta, \phi) \vert q \rangle|^2 }{N_{\mathrm{q}}  ( N_{\mathrm{q}} +1)}
\end{align}
where $\vert q \rangle$ is an orthonormal basis for the qubit states and $N_{\mathrm{q}}$ is the dimensionality of the associated subspace~\cite{pedersen2007fidelity}.
We note that in our implementation we use the reduced symmetric basis described above, with $k = \lfloor N/2 \rfloor$, so the gate fidelity differs slightly from a calculation in the full qubit space. In particular, it weights the symmetric subspace more heavily than lower-spin subspaces. Nevertheless, the basis is complete, so if $\mathcal{F}=1$ in the reduced subspace, it also is unity in the full qubit subspace.

When we optimize dual-driving gates, we further add a regularization to the cost function to bias the solution towards smooth gates. Unlike Refs.~\cite{jandura_time-optimal_2022}, we initialize the optimization with random profiles to avoid local minima, so the smoothening regularization is crucial for obtaining continuous pulse profiles which can realistically be implemented experimentally. 
The combined cost function is therefore
\begin{align}
    C_{\lambda}(\{ \theta, \phi \}) &= ( 1- \mathcal{F}(\{ \theta, \phi \})) \nonumber \\
    + &\lambda \sum_{k} \left[  \left( \frac{\phi(k\tau) - \phi((k-1)\tau)}{2} \right)^2 + \left( \frac{\theta(k\tau) - \theta((k-1)\tau)}{2} \right)^2 \right].
\end{align}
Here, we penalize the derivative of $\phi$ and $\theta$, which correspond to the detuning of the Rydberg drive and amplitude of the hyperfine drive. 

Next, we discuss a systematic procedure for calculating high-fidelity smooth gates (see Extended Data Fig.~\ref{fig:supp_gates_and_higher_order}).
To bias the optimizer towards finding smooth gates, we first minimize the cost function $C_{\lambda}$ with $\lambda = 10^{-3}$, and evaluate each profile by calculating $C_{0}$. For sufficiently long evolution times $T$, we are able to reduce $C_0$ to be about $10^{-3}$, and see the gate is reasonably smooth.
However, the regularization introduces a small amount of bias. As such, we subsequently re-optimize $\theta, \phi$ using $C_0$ as the cost function, and see the ideal fidelity improves dramatically but the final gates remain smooth (see Extended Data~\fref{fig:supp_gates_and_higher_order}). Intuitively, the first step where rapid changes in the pulse profile are penalized confines the subsequent second search into a region of search space where gates are smooth.

\subsubsection{Alternating Ansatz}

A simpler but less time efficient method to realize the required multi-qubit gates involves applying $\Omega_r(t)$ and $\Omega_q(t)$ separately. This alternating approach is a natural extension of the capabilities demonstrated in Ref.~\cite{evered_high-fidelity_2023}, and further generalizes more readily to other reconfigurable architectures~\cite{katz_demonstration_2023,moses2023race}.

We parameterize a generic symmetric diagonal operation, and a global qubit rotation, as
\begin{align}
    D(\phi) &= \exp\left(-i \sum_{j=0}^{N} \phi_j \hat{\mathcal{Z}}_j\right) \label{eqn:diag_multi-qubit}\\ 
    Q(\theta) &= \exp\left(-i \sum_{\alpha = \{x,y,z\}} \theta_{\alpha} \hat{S}^{\alpha}\right)
\end{align}
where $\mathcal{Z}_j = \sum_{i_1,...,i_j} Z_{i_1} Z_{i_2}...Z_{i_j}$ is a linear combination of all weight-$j$ Pauli-$Z$ strings supported on the $N$ qubits.
In particular, a diagonal gate is characterized by time-evolution under the Rydberg part of the Hamiltonian $\Omega_r(t)$, such that all qubit states return to qubit states at the end of the evolution, with a path-dependent phase. Different qubit states acquire different phases, producing a phase gate.
The intuition is that $D(\phi)$ parameterizes all possible diagonal gates which can be generated by the all-to-all Rydberg blockade interaction.
After removing the global phase $\phi_0$, and the part which can be generated by a single-qubit Hamiltonian $\phi_1$, there are $N-1$ remaining parameters which characterize the interaction.
The global qubit rotations $Q(\theta)$ generate single-qubit off-diagonal rotations since $\hat{S}^{\alpha} = \sum_{a=1}^{2S} \hat{s}_a^{\alpha}$, but generate no interactions between qubits.
As such, alternately applying diagonal and off-diagonal operations enables generation of a large class of off-diagonal interaction operations.
To find sequences which realize $U_S$ and $U_P$, we again use GrAPE-style optimization.
In particular, we parameterize the unitary as
\begin{align}
    U_{\mathrm{alternating}} = \prod_{k=1}^{n_{\mathrm{seq}}} Q(\theta_k) D(\phi_k)
\end{align}
and variationally optimize $\theta_k, \phi_k$ to try and match $U_{\mathrm{alternating}}$ to the target unitary.
We identify the shortest sequences which are capable of implementing the target unitary, and present the corresponding sequence lengths ($n_{\mathrm{seq}}$) in Table~\ref{table:alternating_ansatz_nstages}.
\begin{table}[h]
    \centering
    \begin{tabular}{c|c|c|c|c|c|c}
    $n_{\mathrm{seq}}$ & $N=2$ & $N=3$ & $N=4$ & $N=5$ & $N=6$ & $N=7$ \\
    \hline
    $U_S$ &  2 & 3 & 6 & 6 & 6 & 6 \\
    $U_P$ & 2 & 3 & 7 & 7 & 10 & 10
    \end{tabular}
    \caption{Optimized sequence lengths $n_{\mathrm{seq}}$ for target error below ${\sim} 10^{-3}$.}
    \label{table:alternating_ansatz_nstages}
\end{table}
We further note that for $U_S$, the sequences only require using the two-qubit phase $\phi_2$, likely because the evolution Hamiltonian $\hat{\mathbf{S}}^2$ is composed of two-qubit operators. We use this fact below to construct shorter two-qubit decompositions for $U_S$ compared to $U_P$.
Similarly, we find for all sequences ($U_S$ and $U_P$), the odd-qubit phases $\phi_3, \phi_5,...$ can be set to zero.

\begin{figure}
    \centering
    \includegraphics{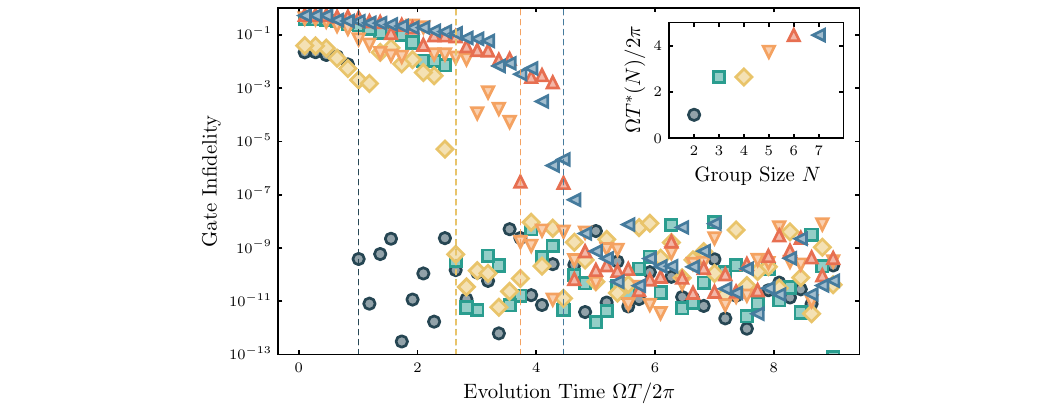}
    \caption{We determine the maximum evolution time $T^*(n)$ required to implement arbitrary symmetric diagonal gates $D(\phi)$ using the blockade interaction for clusters of $N=2,...,7$ qubits. To do this, for each evolution time $\Omega T$ we randomly select a target rotation angles $\phi$, and use GrAPE to find phase profiles implementing $D(\phi)$. Above the critical times $T^*$ (vertical dotted lines), the infidelity achieved was always below $10^{-6}$. Identified values for $T^*(N)$ are also presented in the inset.}
    \label{fig:supp_fig_gates}
\end{figure}

In neutral-atom setups, the global qubit rotation $Q(\theta)$ can be realized with very high fidelity~\cite{levine_parallel_2019}.
As such, the gate errors are dominated by the interaction step $D(\phi)$, which involves coupling to the Rydberg state.
To minimize gate errors, we therefore use GrAPE to identify Rydberg driving profiles $\Omega_r(t)$ that can implement a generic symmetric diagonal gate $D(\phi)$.
For a given system size $N$, by randomly sampling target phases $\phi$, and using GrAPE to optimize the profile independently for each setting, we identify a threshold evolution time $T^*(n)$ above which the GrAPE algorithm always reliably identifies a high-fidelity gate (Fig.~\ref{fig:supp_fig_gates}).
These times, combined with $n_{\mathrm{seq}}$ are used to compute the values in Fig.~\ref{fig:implementation}b.

Finally, we discuss how gates $D(\phi)$ can be conveniently promoted to controlled-$D(\phi)$ gates.
A generic control gate can be parameterized by
\begin{align}
    CD(\phi) = \exp\left(-i \left(\frac{\mathds{1}+Z_a}{2}\right) \sum_{j=0}^{n} \phi_j \mathcal{Z}_j\right)
\end{align}
where $\frac{1+Z_a}{2} = \vert {0} \rangle \langle {0} \vert$ is a projector onto the $\vert {0} \rangle$ state.
We see that $CU_{\mathrm{diag}}$ can be decomposed into two symmetric diagonal gates.
\begin{align}
    U_1 &= \exp\left(\frac{-i}{2} \sum_{j=1}^{n+1} \phi_{j-1} \tilde{\mathcal{Z}}_j \right) \\
    U_2 &= \exp\left(\frac{-i}{2} \sum_{j=1}^{n} (\phi_j - \phi_{j-1}) \mathcal{Z}_j \right)
\end{align}
The first gate applies a symmetric diagonal operation on the ancilla and target qubits, where $\tilde{\mathcal{Z}}_j$ is the linear combination of weight-$j$ Pauli-$Z$ strings supported on the $n+1$ ancilla and target qubits.
This realizes the interactions terms with $Z_a$ on the ancilla site.
However, it also creates additional unwanted terms within the target qubits.
Therefore, the second term simultaneously produces interactions terms with $\mathds{1}$ on the ancilla site, and removes the additional terms from the first step, such that $CU_{\mathrm{diag}} = U_1 U_2$.
This diagonal gate can be combined with the optimized alternating ansatz to construct controlled-spin operations $CU_S$ and $CU_P$. Such operations can be used for controlled probe-state preparation in Fig.~\ref{fig:implementation}a.

Lastly, we note that the MPO construction for the symmetric spin projectors $P_s$, presented in the final section, could be turned into an ancilla-mediated scalable construction for generating spin-gates using two-qubit gates. In particular, the constructed MPO corresponds to a sequence of controlled-gates from the system qubits onto an ancilla register which effectively measures the total-spin information. Then, spin-operations can be systematically generated by applying the MPO circuit between the system and ancilla, applying a diagonal gate on the ancilla, and then undoing the MPO circuit~\cite{lin_lecture_2022}.

\subsection{Heuristic Simulation Time Estimates}

In this section, we provide more details on the heuristics used to estimate coherent simulation times in Fig.~\ref{fig:ham_simulation_comparision}.
The estimates involve breaking down the error contribution into two parts: simulation error and gate error.
Simulation errors arise from imperfect Hamiltonian engineering, due to mismatch between the engineered Hamiltonian $H_F$ and the target Hamiltonian $H_T$.
Gate errors arise from experimental imperfections which accumulate during implementation of few-qubit interactions.
Our goal is to study qualitatively how our Hamiltonian engineering techniques, including both novel Hamiltonian simulation schemes and hardware-optimized multi-qubit operations, can improve the performance of a quantum simulation.

To characterize simulation errors, we look at the decay of the many-body fidelity between the state evolved under $H_F$ and $H_T$.
For symmetrized pulse sequences the decay should exhibit a characteristic $t^2$ and $\tau^4$ dependence.
\begin{align}
    \int d\psi \left[1 - |\langle \psi \vert e^{i H_T t} e^{-i H_F t}\vert \psi \rangle|^2 \right] &= \int d\psi \left[t^2 \left(\langle \psi \vert (H_T - H_F)^2 \vert \psi \rangle - \langle \psi \vert (H_T - H_F) \vert \psi \rangle^2 \right) \right] \\
    &= \int d\psi \left[t^2 \langle \psi \vert (\Delta^2 H_F^{(2)}) \vert \psi \rangle\right] \\
    &\approx (c_2 \tau^2 t)^2
\end{align}
where we have only kept leading contributions in $t$ and $\tau$, averaged over valid initial states $\vert \psi \rangle$, and introduced a coefficient $c_2$ which captures the magnitude of the leading order error term in $H_F - H_T$.
This coefficient depends on the details of the Hamiltonian simulation protocol, and can be understood as arising from the third order term in the Magnus expansion
\begin{align}{\label{eqn:error_coeff}}
    H_F^{(2)} = \frac{\tau^2}{K} \sum_{k_1=1}^K  \sum_{k_2=1}^{k_1} \sum_{k_3=1}^{k_2} \left(1 - \frac{\delta_{k_1,k_2} + \delta_{k_2, k_3} }{2} \right)
    \left(\left[H(k_{1}),\left[H(k_{2}),H(k_{3})\right]\right]+ \left[\left[H(k_{1}),H(k_{2})\right],H(k_{3})\right]\right).
\end{align}

In what follows, we characterize simulation errors via $c_2$, and focus on the effect of the cycle length $K$ on the value of $c_2$.
At a first glance, we observe that in \eqref{eqn:error_coeff} there are ${K \choose 3}$ terms with $k_1 \neq k_2 \neq k_3$, and ${K \choose 2}$ terms with $k_1=k_2$ or $k_2=k_3$, and ${K \choose 1}$ terms where $k_1=k_2=k_3$.
In general, the nested commutators contain a lot of additional structure which will affect $c_2$, including large cancellations between individual terms.
For example, the terms where $k_1=k_2=k_3$ automatically vanish.
Further, for symmetrized pulse sequences, $H(k) = H(K-k+1)$, and many pairs of triple commutators also vanish.
Finally, the fact that each $H(k)$ can be decomposed as $\sum_n H^{(n)} e^{-i \Theta_k n}$, paired with choose pulse sequences such that $\sum_k e^{-i \Theta_k n}$, implies many of the phases will interfere destructively.
A more detailed analysis of the Floquet engineering would take these structures into account explicitly, but is beyond the scope of this work.
Nevertheless, the structure of the third-order term suggests that $c_2$ scales as most as $O(K^2)$, but also may have a large $O(K)$ contribution.
As such, we will utilize numerical simulations to determine a heuristic relationship between $K$ and $c_2$.

To characterize the contribution from incoherent gate errors, we assume the error is accumulated linearly in time.
If we let $g$ be the typical gate error per cycle, then the total incoherent error accumulated during simulation to time $T$ is approximately $gT/\tau$.
We neglect coherent gate errors, which will grow as $T^2$, for simplicity.
Our estimates are based on using hardware-optimized operations for the Rydberg platform, so we estimate $g$ based on the dual-driving gate time from Fig.~\ref{fig:implementation}b, and assume an error rate $g_0 = 10^{-3}$ per physical Rabi time $\Omega T = 2\pi$.

When combined with the Floquet error, sourced by $c_2$, the total error of a simulation can be written as
\begin{align}
    \varepsilon = (c_2 \tau^2)^2 T^2 + \frac{g T}{\tau}
\end{align}
There is a tradeoff between these two error sources, which can be tuned by adjusted the step-size $\tau$.
Minimizing the total error $\varepsilon$ for a fixed evolution time $T$, we see the optimal step-size is
\begin{align}
    \tau_{\mathrm{opt}} = \left(\frac{g}{4Tc_{2}^{2}}\right)^{1/5}
\end{align}
which results in a minimum error of
\begin{align}
    \varepsilon_{\mathrm{opt}} = 5\left(\frac{c^{2}g^{4}T^{6}}{2^{4}}\right)^{1/5}
\end{align}
or equivalently, in a maximum evolution time given a particular error tolerance
\begin{align}
    T_{\mathrm{max}} = \frac{2^{2/3}}{5^{5/6}}\frac{\epsilon^{5/6}}{(cg^2)^{1/3}}
\end{align}

Notice that both $c_2$ and $g$ are extensive quantities, growing linearly in the number of qubits $N$.
As such, we will typically work with intensive quantities, $\tilde{c}_2 = c/N$ and $\tilde{g} = g/N$, and absorb the $N$ dependence into the target error tolerance
\begin{align}
    T_{\mathrm{max}} = \frac{2^{2/3}}{5^{5/6}}\frac{\epsilon^{5/6} / L}{(\tilde{c}_2 \tilde{g}^2)^{1/3}}.
\end{align}
This corresponds to selecting an error which grows (almost) extensively as well, i.e. a (nearly) constant error density.

\begin{figure}
    \centering
    \includegraphics{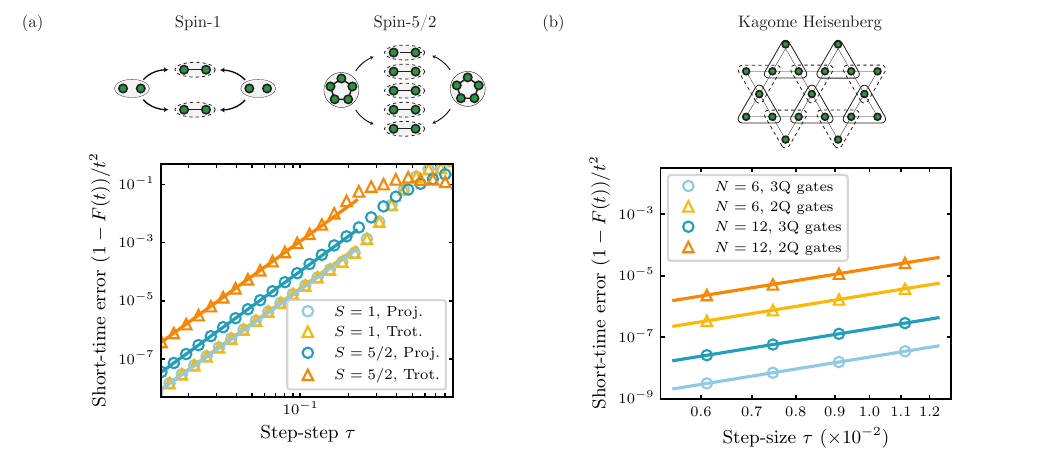}
    \caption{Visualization of procedure used to fit $c_2$ from numerical simulations, for two interacting spins (a) and the Kagome Heisenberg model (b). In (a), we compare a Trotterization approach to one using dynamical projection and multi-qubit gates. For an ensemble of randomly chosen initial states in the symmetric subspace, we time evolve to a target $t$ and compute the average fidelity $F(t) = |\langle \psi \vert U_{\mathrm{ideal}}^{\dagger} U_F(t) \vert \psi \rangle|^2$. We further vary the stepsize $\tau$, and observe the characteristic $\tau^4$ scaling associated with symmetrization, which eliminates the $\tau^2$ term. The overall coefficient $c_2$ is also extracted from the same fits.}
    \label{fig:sim_time_estimates}
\end{figure}

\subsubsection{Two interacting spin-$S$'s}

We next describe how $c_2$ and $g$ are estimated heuristically for the four models studied in Fig.~\ref{fig:ham_simulation_comparision}.
We start by simulating the two spin-$S$ model with generic two-body interactions described by $H_2 = \sum_{\alpha \beta} J^{\alpha \beta} \hat{S}^{\alpha}_1 \hat{S}^{\beta}_2$.
In particular, we consider a mixture of Heisenberg and DM interactions, $J^{\alpha \beta} = J \delta^{\alpha \beta} + D \sum_{\gamma} \epsilon^{\alpha \beta \gamma}$.
The $c_2$ coefficient is fitted numerically from noise-free dynamics, for both a two-qubit Trotter and multi-qubit dynamical projection approach. 
In particular, we simulate ideal and simulated Hamiltonian dynamics from random product states $\vert \psi \rangle$ in the symmetric subspace, and compute the error at short times $t$.
Then, we extract the coefficient $c_2 \tau^2$ by averaging over a small time window, and fit $c_2$ by varying $\tau$ (see Fig.~\ref{fig:sim_time_estimates}).

In the Trotter approach, the Hamiltonian simulation protocol involves applying a sequence of $2S$ Hamiltonians, each of which act pairwise between the two clusters.
\begin{align}
    (H_{2, \mathrm{Trotter}})_i = \sum_{\alpha \beta} \sum_{a=1}^{2S} \bar{J}_{ij}^{\alpha \beta} \hat{s}_{i,a}^{\alpha} \hat{s}_{j,a+i \, \mathrm{mod}({2S})}^{\beta}
\end{align}
This decomposition ensures every pair of constituent qubits between the two clusters interacts exactly once during the $2S$ step-sequence. 
Note that we say this sequence has $K = 2S$, even though the symmetrized sequence we simulate actually has period $2K$.

In the dynamical projection approach, we select $H_{2, I}  = (H_{2, \mathrm{Trotter}})_0$ to be the first pairing of the cluster of qubits.
It turns out, for this problem, a $K=2$ sequence is sufficient to symmetrize $H_{2,I}$. 
The key observation, is to notice that $H_{2,I}^{(\pm 1)} = 0$, due to the specific structure of the Hamiltonian.
Decomposing into all terms we get,
\begin{align}
    H_{2,I}(k) &= \sum_{a=1}^{2S} \bar{J}^{\alpha \beta} \big({\hat{s}_{1,a}^{\alpha}}^{(0)} {\hat{s}_{2,a}^{\beta}}^{(0)} \nonumber\\
    &+ {\hat{s}_{1,a}^{\alpha}}^{(+1)} {\hat{s}_{2,a}^{\beta}}^{(0)} e^{i \Theta_k} + {\hat{s}_{1,a}^{\alpha}}^{(0)} {\hat{s}_{2,a}^{\beta}}^{(+1)}e^{i \Theta_k} \nonumber\\
    &+ {\hat{s}_{1,a}^{\alpha}}^{(-1)} {\hat{s}_{2,a}^{\beta}}^{(0)} e^{-i \Theta_k} + {\hat{s}_{1,a}^{\alpha}}^{(0)} {\hat{s}_{2,a}^{\beta}}^{(-1)}e^{-i \Theta_k} \nonumber\\
    &+ {\hat{s}_{1,a}^{\alpha}}^{(1)} {\hat{s}_{2,a}^{\beta}}^{(1)} e^{2 i \Theta_k} + {\hat{s}_{1,a}^{\alpha}}^{(-1)} {\hat{s}_{2,a}^{\beta}}^{(-1)}e^{-2i \Theta_k} \big).
\end{align}
However, notice that ${\hat{s}_{j,a}^{\beta}}^{(0)} = \frac{1}{2S} S_{j}$, and $\sum_a {\hat{s}_{1,a}^{\alpha}}^{(\pm n)} = 0$ for any $n \neq 0$ due to symmetry.
As such, the terms which change $H_P$ by $\pm 1$ vanish, leaving only the $\pm 2$ terms.
This means the cancellation condition can be satisfied using the two-step sequence $\Theta_k = (0, \pi/2)$.

The fitted values of $c_2$ from numerical simulations are presented in Table.~\ref{table:twospin_c2_scaling}.
In particular, we present an intensive version $\tilde{c_2}$ which is normalized by the total number of qubits in the system.
Furthermore, note that the value of $c_2$ depends on the characteristic energy scale of the Hamiltonian.
To handle this systematically, we scale the simulation time by a simple measure of the local Hamiltonian strength $|H|_{\mathrm{local}}$.
This is computed by simply adding the number of Pauli operators which intersect with any given qubit $i$ weighted by the magnitude of their coefficients, and dividing by the weight of each operator.
For example, for $H_2$, there are $2S \times 3$ two-body Pauli operators with coefficient $J$, and $2S \times 6$ two-body operators with coefficient $D$. Therefore, $|H|_{\mathrm{local}} = 6S ( J + 2D) / 2= 9SJ$ for $D = J$.

\begin{table}[h]
    \centering
    \begin{tabular}{c|c|c|c|c}
    $\tilde{c}_2$ & $S=1$ & $S=3/2$ & $S=2$ & $S=5/2$ \\
    \hline
    Trotter &  0.135 & 0.215 & 0.285 & 0.397 \\
    Dynamical Projection & 0.12 & 0.17 & 0.17 & 0.18
    \end{tabular}
    \caption{Numerically estimated values for the normalized, intensive coefficients $\tilde{c}_2$, for two interacting large spin $S$'s. The Trotter error is consistent with roughly linear growth, arising the growth of the sequence length $K = 2S$. In contrast, dynamical projection appears consistent with a constant $\tilde{c}_2$, matching the constant sequence length $K=2$.}
    \label{table:twospin_c2_scaling}
\end{table}

The computation of $g$ involves counting the total number of gate per step. In the Trotter approach, there are $2S$ two-qubit interactions. 
Because each interaction involves a symmetric and anti-symmetric part, it needs to be further decomposed, e.g. into a sequence of symmetric gates and single-qubit rotations.
In particular, a three step sequence of symmetric interactions $h_1 = J (s_1^x s_2^x + s_1^y s_2^y + s_1^z s_2^z)$, $h_2 = D(s_1^x s_2^y + s_1^y s_2^z + s_1^z s_2^x)$, and $h_3 = D(s_1^x s_2^z + s_1^y s_2^x + s_1^z s_2^y)$ can realize the target Hamiltonian on average. Importantly, $h_2$ and $h_3$ can be brought into the same form as $h_1$ by applying a single-qubit rotation on the second site around the $s^x + s^y + s^z$ axis. Note that, while this introduces additional simulation error at higher-order, variational gate optimization as in the alternating ansatz could be used to identify higher-fidelity local sequences.
As such, each cycle in the Trotter approach utilizes $2S \times 3$ two-qubit symmetric gates.
In the dynamical projection approach, there are still $2S \times 3$ two-qubit gates per cycle, but also two additional $2S$-qubit gates to apply $H_P$.
However, in practice some pairs of gates applied during separate steps can be merged, such as when the interaction frame $\Theta_k = \Theta_{k+1}$ is the same.
In symmetrized sequences, this happens during the $1$st and $K$-th gates in the $2K$ cycle.
As such, we multiply $g$ by $\frac{K-1}{K}$ to account for this reduction.
The intensive quantity $\tilde{g}$ is presented in Table.~\ref{table:twospin_g_scaling}.

\begin{table}[h]
    \centering
    \begin{tabular}{c|c|c|c|c}
    Gate error $\tilde{g}$ & $S=1$ & $S=3/2$ & $S=2$ & $S=5/2$ \\
    \hline
    Trotter &  3.1 & 4.1 & 4.7 & 5.0 \\
    Dynamical Projection & 4.1 & 4.2 & 4.0 & 3.9
    \end{tabular}
    \caption{Numerically estimated values for the gate error per cycle $(\times 10^{-3})$, per qubit, $\tilde{g}$. Note that despite the overhead, the projection approach becomes more efficient than Trotter due to the ability to combine gates during the $1$st and $K$-th cycle. Further, the projection overhead decreases with $2S$ since this quantity is intensive.}
    \label{table:twospin_g_scaling}
\end{table}

Since $K$ is constant for dynamical projection, and grows with $2S$ for Trotterization, the dynamical projection approach becomes more favorable as the spin-size grows.
Combining the estimated values for $c_2$ and $g$ produces the simulation times in Table~\ref{table:twospin_times_scaling}.

\begin{table}[h]
    \centering
    \begin{tabular}{c|c|c|c|c}
    Sim. Times ($\times |H|_{\mathrm{local}}$) & $S=1$ & $S=3/2$ & $S=2$ & $S=5/2$ \\
    \hline
    Trotter &  5.6 & 3.9 & 3.3 & 2.8 \\
    Dynamical Projection & 4.8 & 4.2 & 4.3 & 4.3
    \end{tabular}
    \caption{Numerically estimated simulation times, for an (approximately) per-qubit error tolerance of $\epsilon = 0.1$, in units of $|H|_{\mathrm{local}}$.
    }
    \label{table:twospin_times_scaling}
\end{table}

\subsubsection{Kagome Heisenberg}

Next, we study the Kagome Heisenberg model.
Since this model has no large-spins, there is no need for the projection approach. However we can still utilize multi-qubit gates to improve the simulation.
To illustrate this, we consider two cases, one with two-qubit gates and another with three-qubit gates.
In the two-qubit case, we consider a $K=6$ protocol, where two-qubit $\hat{\mathbf{s}}_i \cdot \hat{\mathbf{s}}_j$ gates are applied during each step.
In the three-qubit case, we develop a $K=2$ protocol, applying three-qubit $\hat{\mathbf{S}}^2$ evolution along upwards-facing triangles and downwards-facing triangles in alternating fashion.

We follow the same fitting procedure for $c_2$ with $N=6$ and $N=12$ size systems with periodic boundary conditions, and again estimate $g$ using the gate-times from Fig.~\ref{fig:implementation}b.
The $N=12$ values are used for the estimates in Fig.~\ref{fig:ham_simulation_comparision}.

\begin{table}[h]
    \centering
    \begin{tabular}{|c|c|c|c|}
        \hline
        \textbf{$N$} & \textbf{$\tilde{c}_2$} & \textbf{$\tilde{g}$} & \textbf{$T_{\mathrm{max}}$} \\
        \hline
        $N=6$, two-qubit gates & 2.6 & 1.7 & 3.1 \\
        $N=6$, three-qubit gates & 0.25 & 1.7 & 6.9 \\
        $N=12$, two-qubit gates & 3.5 & 1.7 & 2.8 \\
        $N=12$, three-qubit gates & 0.36 & 1.7 & 6.1 \\
        \hline
    \end{tabular}
    \caption{Parameters used to estimate effective simulation time for the Kagome Heisenberg}
    \label{table:kagome_heisenberg}
\end{table}

We notice that the $\tilde{c}_2$ coefficients are signficantly higher for the $K=6$ protocol compared to the similar $K=5$ spin-5/2 value. This discrepancy could be partly due to the fact that the local Hamiltonian norm is overestimated in the two spin-5/2 case, which has more frustrated (non-commuting) interaction terms. In particular, under a time rescale $t \rightarrow \lambda t$, the $c_2$ coefficient varies by $\lambda^3$, making its specific value quite sensitive to the choice of units.
Nevertheless, the overall trend is clear, shortening the sequence appears to accumulate simulation errors more slowly. 

\subsection{Spin-2 models}

Finally, we use this approach to present heuristic estimates of the simulation time for a representative complex spin-model.
In particular, we consider a system of spin-2's on a square lattice, with pairwise higher-order terms $(\mathbf{S}_i \cdot \mathbf{S}_j)^k$, making the interactions quite complicated.
\begin{align}
    H_{\mathrm{sq},n} = \sum_{k=1}^{n} \sum_{\langle i j \rangle} \frac{J_n}{(2S)^{n-1}} (\mathbf{S}_i \cdot \mathbf{S}_j)^k
\end{align}

One of the key limitations towards realizing this model with the conventional two-qubit Trotter approach stems from the fact that each term $(\mathbf{S}_i \cdot \mathbf{S}_j)^k$ decomposes into a sum of $2k$-qubit interactions. Further, decomposing higher-order interactions into two-qubit gates comes with large overhead, as depicted in Fig.~\ref{fig:implementation}c.
The second is the the rapid growth of the number of terms with the order of the interaction. For example, since each $S=2$ site has four qubits, there are ${4 \choose 2}^2 = 36$ groups of four-qubit interactions coupling adjacent sites $i$ and $j$ in $(\mathbf{S}_i \cdot \mathbf{S}_j)^2$.
Both of these problems are mitigated in the multi-qubit approach. In particular, the hardware-level optimization introduces native multi-qubit gates to directly realize higher-qubit operations.
Further, dynamical projection realizes the multitude of terms using simpler operations.

We first consider the $H_2$ model with Heisenberg $k=1$ and bi-quadratic $k=2$ interactions.
The Trotter version involves cycling through the $144$ four-qubit interactions, between each spin and its four neighbors. Since each interaction acts on two-qubits per cluster, this can be parallelized into a cycle of period $K=72$. 
In contrast, the dynamical projection approach alternates between vertical interactions, horizontal interactions, and an intra-cluster interaction ($H_P$). Since $n_{max}=2$, this needs to be cycled three times, leading to $K=6$.
In order to get a heuristic estimate for $c_2$, we extrapolate a linear-scaling, and simply approximate $\tilde{c}_2 \sim K$, which is within an order-of-magnitude of the estimates from simulation above.

Decomposition of the four-qubit spin interaction into two-qubit operations requires roughly 100 gates, resulting in a gate-error per qubit per step of $100 g_0 / 4 = 25 g_0$. In contrast, the large-angle four-qubit simultaneous driving gate takes time $3.74 g_0$.
Since there are two interaction gates for every intra-cluster gate, this results in a gate-error per qubit per step of $3.74*1.5 g_0 / 4 = 1.395 g_0$. Note for simplicity, we neglect the additional $\frac{K-1}{K}$ factor arising from combining gates.

Putting this together, our estimated simulation time is $T_{\mathrm{max}} = 0.17$ for the conventional, two-qubit Trotterization approach, and $T_{\mathrm{max}} = 2.7$ for the multi-qubit dynamical projection approach.
We can also isolate the relative contribution from multi-qubit gates and dynamical projection.
The multi-qubit Trotter approach, (using $c_2=77$ and $g = 1.395 g_0$) is estimated to be $T_{\mathrm{max}} = 1.2$.
The two-qubit, dynamical projection approach, (using $c_2=77$ and $g = 1.395 g_0$) is estimated to have $T_{\mathrm{max}} = 0.4$.

Finally, we consider a model with up to bi-quartic $k=4$ interactions. Here, the $k=4$ terms act on all eight qubits between two clusters, so no dynamical projection is needed. Instead, all of the advantage comes from the efficiency of the multi-qubit approach.
The trotterization period is $K=4$, and the two-qubit decomposition estimates suggests over $10^4$ gates are needed to realize the $k=4$ interaction, so we plug in $g \sim 10^4 g_0 / 8 \sim 10^3 g_0$.
In contrast, the eight-qubit gate times is $\Omega T / 2\pi = 5.0$, producing $g = 5.0 g_0 / 8$.
Again using $c_2 = K$ heuristically, this results in estimated evolution times of $0.03$ and $5.3$ respectively.

\section{Many-body Spectroscopy}
The algorithmic framework we develop utilizes classical co-processing to extract detailed spectral information from time-evolution and snapshot measurements in a resource-efficient way.
In the following sections, we discuss the following key aspects.
First, we describe how measurements from quantum simulation (Fig.~\ref{fig:spectroscopy_toy_model}a) enable parallel measurement of exponentially many two-time correlation functions of the form $C_{O,R}(t)$. 
Next, we explain how from $M$ snapshots, and for simple reference states $\vert S\rangle$, this information is in principle sufficient to efficiently reconstruct \emph{any linear functional} of the time-evolved probe states $U(t) \vert R \rangle$.
We show how this enables efficient calculation of the density-of-states, and how the convergence of the estimator can be system size independent for appropriately chosen ensembles.
Then, we discuss the estimation of thermal expectation values, and finally present a $\mathrm{poly}(N)$ MPO construction for computing spin-projected DOS, enabling efficient classical evaluation of $D^{P_S}(\omega)$.

\subsection{Estimator \eqref{eqn:COR_estimator} for $C_{O,R}(t)$ from snapshots}

The core quantities which enable access to spectral properties of the Hamiltonian are correlation functions of the form $C_{O,R}(t) = \langle S \vert O(t) R(0) \vert S \rangle$,
For a given evolution time $t$ and perturbation $R$, these can be efficiently obtained using the circuit in Fig.~\ref{fig:spectroscopy_toy_model}a. 
Further assume the system qubits are measured in a fixed basis, such as the $X$-basis, and $O$ is diagonal in this basis.
Then, $C_{O,R}$ can be efficiently estimated in parallel from the resulting snapshot measurements.
During the $i$-th run of the experiment, let $\mu^{(i)} = \{ x,y \}$ be the randomly selected ancilla measurement basis each with probability 1/2, and $a^{(i)} = \{0,1\}$, and $\vert b^{(i)} \rangle$ be the ancilla and system measurement outcomes respectively.
Then, as presented in \eqref{eqn:COR_estimator}, the estimator can be written as
\begin{align}
    \overline{C_{O,R}(t)} = \frac{1}{M}\sum_{i=1}^{M} 2\sigma(\mu^{(i)}, a^{(i)}) \langle b^{(i)} \vert O \vert b^{(i)} \rangle.
\end{align}

To study the performance of this estimator, first we show that it is unbiased, in the sense that it's expectation value matches the target correlation function. Then, we compute it's variance, which determines how quickly the estimator converges to it's expectation value. 

To show the estimator is unbiased, we need to compute the probability that during any given run, the output of the simulation is $\mu, a, b$.
The conditional measurement probabilities $P(a,b | \mu)$ can be straightforwardly computed using the form of of $\vert \psi_f \rangle$.
\begin{align}
    P(0,b | x) = \frac{1}{2} \langle \mathds{1} \otimes \Pi_b \rangle_{\psi_f} + \frac{1}{2} \langle X \otimes \Pi_b \rangle_{\psi_f} \\
    P(1,b | x) = \frac{1}{2} \langle \mathds{1} \otimes \Pi_b \rangle_{\psi_f} - \frac{1}{2} \langle X \otimes \Pi_b \rangle_{\psi_f} \\
    P(0,b | y) = \frac{1}{2} \langle \mathds{1} \otimes \Pi_b \rangle_{\psi_f} + \frac{1}{2} \langle Y \otimes \Pi_b \rangle_{\psi_f} \\
    P(1,b | y) = \frac{1}{2} \langle \mathds{1} \otimes \Pi_b \rangle_{\psi_f} + \frac{1}{2} \langle Y \otimes \Pi_b \rangle_{\psi_f}
\end{align}
where $\Pi_b = \vert b \rangle \langle b \vert$ is the projector onto the measured state.
For observables $O$ which are diagonal in the $X$-basis, it can be uniquely decomposed in the operator basis spanned by the projectors $\Pi_b$,
\begin{align}
    O = \sum_b \Pi_b \langle b \vert O \vert b \rangle.
\end{align}
Therefore, we see that the expectation value of the estimator is equal to
\begin{align}
    \mathds{E}\left[ 2\sigma(\mu,a)\langle b \vert O \vert b \rangle \right] &= \sum_{a} \sum_{b} \left(P(a,b | x)P(x)2\sigma(x,a) + P(a,b | y) P(y)2 \sigma(y,a) \right)\langle b \vert O \vert b \rangle \\
    &= \left(\frac{1}{2}\langle \mathds{1} \otimes \Pi_b \rangle_{\psi_f} + \frac{1}{2} \langle X \otimes O \rangle_{\psi_f}\right) - \left(\frac{1}{2}\langle \mathds{1} \otimes \Pi_b \rangle_{\psi_f} - \frac{1}{2} \langle X \otimes O \rangle_{\psi_f} \right) \\
    &+i\left(\frac{1}{2}\langle \mathds{1} \otimes \Pi_b \rangle_{\psi_f} + \frac{1}{2} \langle Y \otimes O \rangle_{\psi_f}\right) -i\left(\frac{1}{2}\langle \mathds{1} \otimes \Pi_b \rangle_{\psi_f} - \frac{1}{2} \langle Y \otimes O \rangle_{\psi_f}\right) \\
    &= \langle (X + iY) \otimes O \rangle_{\psi_f} = C_{O,R}
\end{align}
as desired.
Note that the factor of two in the coefficient $2\sigma(\mu,a)$ accounts for the fact that each basis is only sampled half the time, cancelling the contribution from $P(\mu)$.

The variance of the estimator can be computed by performing similar calculations.
Formally, the second moment of the estimator is given
\begin{align}
    \mathds{E}\left[ |2\sigma(\mu,a)\langle b \vert O \vert b \rangle|^2 \right] &= \mathds{E}\left[ 4|\langle b \vert O \vert b \rangle|^2 \right] \\
    &= 4 \sum_b \langle \Pi_b \rangle_{\psi_f} |\langle b \vert O \vert b \rangle|^2 \\
    &= 2 \sum_b (\langle R(t) \vert b \rangle \langle b \vert R(t) \rangle+\braket{S(t)|b}\braket{b|S(t)})|\langle b \vert O \vert b \rangle|^2 
\end{align}
Here, we used the fact that $|\sigma(\mu,a)|^2=1$ for all choices of $\mu$ and $a$; thus, the properties of the ancilla drop out of the expression. Finally, subtracting the expected value gives an expression for the variance.

\subsubsection{Beyond diagonal operators}

When the operator $O$ is not necessarily diagonal in the measurement basis, estimating $C_{O,R}(t)$ will require varying the measurement basis of the system qubits as well.
State-of-the-art shadow tomography methods could be utilized for this purpose~\cite{huang_predicting_2020,elben_randomized_2022}.
We discuss here the simplest case, random Pauli measurements, where the joint ancilla and system measurement basis is $\mu \nu = \mu \nu_1 ... \nu_n$. As before, $\mu$ determines whether the ancilla is measured in $X$ or $Y$, while $\nu_1,...,\nu_n$ determines whether each system qubit is measured in $X$, $Y$, or $Z$.
An operator $O$ can be decomposed into a linear combination of Pauli operators, $O = \sum_{P} O(P) P$, where $P = P_1 ... P_n$ is a product of Pauli operators, and $O(P) = \mathrm{tr}(O P)$.
Furthermore, define $\vert b^{(i)} \rangle$ as the system state measured during the $i$-th run, which depends on the choice of basis setting $\nu^{(i)}$ and measured bitstring. 
Then, the estimator $\overline{C_{O,R}(t)}$ can be constructed as
\begin{align}{\label{eqn:COR_pauli}}
    \overline{C_{O,R}(t)} \simeq \frac{1}{M}\sum_{i=1}^{M} 2  \sigma(\mu^{(i)}, a^{(i)}) \sum_{P} 3^{|P|} O(P)  \langle b^{(i)} \vert P \vert b^{(i)} \rangle
\end{align}
The additional factor of $3^{|P|}$ comes from inverting the probability that the measurement basis is aligned with the Pauli operator $P$, where $|P|$ is the number of sites non-identity sites~\cite{huang_predicting_2020}.
Indeed, if the measurement basis is not aligned, then $\langle b^{(i)} \vert P \vert b^{(i)} \rangle = 0$ instead of $\pm 1$, and so no information about the expectation value of $P$ is learned.

\subsection{Interpretation as shadow tomography of $\vert R(t) \rangle$ from single-basis measurements}

Snapshot measurements enable construction of a classical representation, i.e., a classical shadow~\cite{huang_predicting_2020}, of $\vert R(t) \rangle\,{=}\,U(t) R \vert S \rangle$, from which linear functionals (such as wavefunction amplitudes $\langle \phi \vert R(t) \rangle$) can be efficiently estimated. The amplitude used for the DOS calculation $\langle R \vert R(t) \rangle$ is just one such example. 

To illustrate this formalism, let us still again consider the case where $\vert S \rangle = \vert {0} \rangle^{\otimes N}$, and snapshot measurements of the system are performed in the $X$-basis, where all Pauli strings are enumerated as $X_s$. Then, $\{ X_s \ket{0}^{\otimes N}, s\in \{0,...,2^N-1\} \}$ forms a complete, orthonormal basis for the Hilbert space and we can again utilize the resolution of the identity in terms of $X_s$ acting on a classical state, 
\begin{equation}
\mathds{1} = \sum_s X_s \vert {0}^N \rangle \langle {0}^N \vert X_s,
\end{equation}
to rewrite $\vert R(t)\rangle$ as a linear combination correlation functions $C_{X_s,R}(t)$, which can be measured simultaneously for each $s$, 
\begin{align}{\label{eqn:R(t)_expression}}
    \vert R(t) \rangle &= \sum_s X_s \vert {0}^N \rangle \langle {0}^N \vert X_s \vert R(t) \rangle \nonumber \\
    &= \sum_s X_s \vert {0}^N \rangle C_{X_s,R}(t) e^{-i E_{0} t},
\end{align}
where we additionally use the fact that $\ket{0}^{\otimes N}$ is an eigenstate of the Hamiltonian with energy $E_{0}$ and does not evolve in time. We can use the estimator of $C_{X_s,R}(t)$ from Eq.~\eqref{eqn:COR_estimator} to form an estimator for linear functionals of $\vert R(t) \rangle$, which makes Eq.~\eqref{eqn:R(t)_expression} analagous to a classical shadow. 
The classical shadow is an object in Ref.~\cite{huang_predicting_2020}, composed of snapshot measurements, which replicates the properties of a density matrix $\rho$ as the number of measurements $M \rightarrow \infty$ tends to infinity. 
The verison here, tends to the wavefunction $\vert R(t) \rangle$, and is hence estimators for wavefunction amplitudes $\langle \phi \vert R(t) \rangle$ can be written as
\begin{align} \label{eq:phi_R_estimator}
    \overline{\langle \phi \vert R(t) \rangle} &= \frac{e^{-i E_{0} t}}{M}\sum_{i=1}^{M} 2 \sigma(\mu^{(i)}, a^{(i)})  2^{N/2} \langle \phi \vert b^{(i)} \rangle
\end{align}
where $\vert b^{(i)} \rangle$ is the $i$-th measured $X$-basis bitstring. To be able to efficiently evaluate this expression, it is important that $\vert \phi \rangle$ has an efficient classical representation, so that the coefficient $\langle \phi \vert b^{(i)} \rangle$ are computable classically.
Note that, while any wavefunction amplitude can be efficiently computed. 

Unlike the original versions of shadow tomography~\cite{huang_predicting_2020}, here we utilize partial knowledge of the measured state (namely the form of $\vert S \rangle$) in order to extract all of the unknown information (functionals of $\vert R(t)\rangle$) from a single-basis measurement.
Further, our procedure is sensitive to global phase information in $\vert R(t)\rangle$, since it works by performing interferometry relative to a reference state $\vert S \rangle$. 
This phase information is crucial for estimation of spectral information via the Fourier-transformed state $\vert R(\omega)\rangle$.
In Ref.~\cite{chan_algorithmic_2022}, snapshot measurements were combined with time-series data, to extract a different Hamiltonian spectrum that has peaks corresponding to energy differences $\epsilon_i - \epsilon_j$ between eigenstates.
Our method, utilizing the ancilla and reference state, is capable of detecting individual energies $\epsilon_i$ and filtering their corresponding eigenstates~\cite{lu_algorithms_2021}.

\subsection{The operator-resolved DOS estimator}

The key quantity required to evaluate the DOS estimator \eqref{eq:dos_detailed} is the calculation of $D^{A}_R$ defined in \eqref{eqn:DAR_exact}.
One way of arriving at the expression for the estimator, is by noticing that $D^{A}_R = \langle R \vert A \vert R(t) \rangle$, which is equivalent to selecting $\vert \phi \rangle = A \vert R \rangle$ in \eqref{eqn:R(t)_expression}.
This estimator is also equivalent to computing $C_{O,R}$ for a non-hermitian operator $O$.
Focusing for simplicity on the polarized reference state $\vert S \rangle = \vert 0 \rangle^{\otimes N} = \vert 0^N \rangle$, the observable is
\begin{align}
    O:=O^A_R = \sum_s \langle R \vert A X_s \vert 0^N \rangle \hat{X}_s,
\end{align}
and can be written as a linear combination of Pauli-$X$ operators. We can verify that $C_{O_R^A,R}$ reproduces $D_R^A$  up to the global time-dependent phase shift,
\begin{align}
    C_{O^A_R,R}(t) &= \sum_s \langle R \vert A X_s \vert 0^N \rangle \langle 0^N \vert \hat{X}_s(t) \vert R \rangle = e^{i E_0 t}D_R^A(t),
\end{align}
where we assumed that $\ket{0^N}$ is an eigenstate of $H$ with energy $E_S$. 

Despite the exponentially large sum over $\{s\}$, the density-of-states estimator can often be evaluated efficiently from snapshot data, as long as the states $\ket{R}$ and $\ket{S}$ are chosen correctly.
For the polarized reference state the diagonal matrix elements take a simple form,
\begin{align}{\label{eqn:ORA_estimator}}
    \langle b \vert O_R^A \vert b \rangle &= \sum_s \langle R \vert A X_s \vert {0}^N \rangle \langle b \vert \hat{X}_s \vert b \rangle \nonumber\\
    &= \langle R \vert A \sum_s (-1)^{b \cdot s} X_s \vert {0}^N \rangle \nonumber\\
    &= 2^{N/2}\,\langle R \vert A  \vert b \rangle,
\end{align}
which is straightforward to evaluate for many choices of the probe state $\ket{R}$ and operators $A$.
A large class of states for which this can be efficiently evaluated classically, is when $\vert R \rangle, \vert S \rangle$ are matrix product states (MPS) with $\mathrm{poly}(N)$ bond-dimension, and $A$ is an matrix product operator (MPO) with $\mathrm{poly}(N)$ bond-dimension.
For example, the projectors onto different total spin-sectors $P_s$ we leverage in Fig.~\ref{fig:spectroscopy_toy_model} and Fig.~\ref{fig:OEC} can be written efficiently using the MPO formalism, as we discuss further below.

\subsubsection{Convergence of the DOS estimator}

To compute how quickly the DOS ($A\,{=}\,\mathds{1}$) calculation will converge for polarized reference $\vert S \rangle = \vert 0 \rangle^{\otimes N}$ and $O_s = X_s$, we bound the variance of the estimator, and discuss the dependence on the ensemble of probe states.
For simplicity, let us focus on $t=0$, where $C_{O_R,R}(0) = 1$ by construction.
The second moment is equal to
\begin{align}
    \mathds{E} |\overline{C_{O_R,R}}|^2 &= 2^{N+1} \sum_b (\abs{\langle R \vert b \rangle}^2 + \abs{\langle S \vert b \rangle}^2)\abs{\langle R \vert b \rangle}^2=2+2^{N+1}\sum_b \abs{\langle R \vert b \rangle}^4,
\end{align}
where we used $\abs{\braket{S|b}}^2\,{=}\,2^{-N}$. We evaluate this expression in a few relevant limits.
First, we consider states $\vert R\rangle$ which come from applying random single-qubit rotations of the form $\prod_i e^{-i \eta_i X_i}$, where $\eta_i$ encodes the random rotation angles.
For this choice, it is straightforward to show that $\langle R \vert b \rangle \langle b \vert R \rangle = 2^{-N}$ for every $X$-basis eigenstate $\vert b \rangle$. Intuitively, each qubit of $\vert R \rangle$ lives in the $YZ$-plane, and hence is orthogonal to the $X$-axis.
Then, the second moment evaluates to
\begin{align}
     \mathds{E} |\overline{C_{O_R,R}}|^2 &= 2+ 2^{N+1} \sum_b \abs{\langle R \vert b \rangle}^4\to 2+2^{N+1}\times2^{N}\times 2^{-2N} = 4,
\end{align}
and the variance is ${\rm Var}(\overline{C_{O_R,R}})\,{=}\,\mathds{E} [\abs{\overline{C_{O_R,R}}}^2]\,{-}\,\abs{\mathds{E} [\overline{C_{O_R,R}}]}^2\,{=}\,3$.
More generally, states $\vert R \rangle$ generated by evolving under a Hamiltonian diagonal in the $X$ basis will satisfy these results.

Next, we consider another limiting case, whereby $\vert R \rangle$ is an $N$-qubit Haar random state. The Haar-averaged second moment can be written as,
\begin{equation}
    \mathds{E}_{R\in{\rm Haar}} \mathds{E} |\overline{C_{O_R,R}}|^2 = 2+2^{N+1}\sum_b\int dR\,\abs{\braket{R|(\ket{b}\bra{b})|R}}^2,
\end{equation}
which is expressed via a random-state average of the $\ket{b}\bra{b}$ operator. We use the formula for Haar-random averages (Eq.~(1) in Ref.~\cite{pedersen2007fidelity}) with $M\,{=}\,\ket{b}\bra{b}$,
\begin{align}
\int dR\, |\langle R \vert M \vert R \rangle |^2  = \frac{1}{2^N(2^N+1)} \left[ \mathrm{Tr}(M M^{\dagger}) + |\mathrm{Tr}(M)|^2 \right],
\end{align}
which results in the variance,
\begin{align}
    {\rm Var}(\overline{C_{O_R,R}})_{\rm Haar} &= 2+2^{N+1}\times2^N\times \frac{1+1}{2^N(2^N+1)} - 1 = 1 + \frac{4}{1+2^{-N}},
\end{align}
which is of order 1 for large systems. Moreover, the above formula has a product structure, which allows us to also analyze the case where the perturbation $R$ is a product of locally Haar-random operations. Concretely, we can divide $N$ qubits into $n$ groups of $k\,{=}\,N{/}n$ qubits and assume that $\vert R \rangle$ is a 2-design within each group (e.g. is a Clifford circuit).
Then, we get
\begin{align}
	{\rm Var}(\overline{C_{O_R,R}})_{\rm local\,Haar} &= 2+2\prod_{i=1}^{n}  2^k \sum_{b_i} |\langle R_i \vert b_i \rangle \langle b_i \vert R_i \rangle|^2 -1 = 1+2\left( \frac{2}{1+2^{-k}} \right)^n,
\end{align}
which reproduces the previous result for $n\,{=}\,1$. Now, if we consider a fixed subsystem size $k$ that is independent of $N$, the variance will grow exponentially with $N$. However, if the subsystem size increases with $N$ (fixed number of partitions $\alpha$), such that $k\,{=}\,\alpha N$,  then the expression will saturate to a constant $2^{\alpha}$. Thus, the variance is larger than for $k\,{=}\,N$, but still independent of the system size. Since variance is directly related to sample complexity, it is important to avoid exponentially growing variances in practical settings.
We note that maintaining low variance while considering optimized state-preparation strategies, such as MPS state preparation, is an interesting problem we leave for future work.

\subsubsection{More general reference states}

Spectroscopy from more general reference states $\vert S \rangle$ can also be performed, by considering different measurement strategies. One approach is to apply the inverse state-preparation unitary $S^{\dagger}$ and then measure in the $X$-basis.
This is enables parallel access to the $2^N$ operators $\{ S X_s S^{\dagger} \}$, which by construction form an orthonormal basis when applied to the reference state $\vert S \rangle$.
In particular, we can use this ensemble to construct an alternate resolution of the identity
\begin{align}
\mathds{1} &= \sum_s S X_s S^{\dagger} \vert S \rangle \langle S \vert S X_s S^{\dagger} = S \left( \sum_s X_s \vert 0^N \rangle \langle 0^N \vert X_s \right) S^{\dagger} = S S^{\dagger},
\end{align}
which can be used as before to compute $D^{A}_R(t)$, 
\begin{align}
    D^{A}_R(t) &= \sum_s \langle R \vert A S X_s S^{\dagger} \vert S \rangle \langle S \vert S X_s S^{\dagger} \vert R(t) \rangle \nonumber \\
    &= \sum_s \langle R \vert A S X_s \vert {0}^N \rangle C_{S X_s S^{\dagger},R}(t) e^{-i E_{0} t}.
\end{align}
As with \eqref{eqn:ORA_estimator}, this sum can be efficiently evaluated classically as long as the state-preparation unitary $S$ has an efficient MPO description.

Randomized measurements also provide an alternate route for extracting the density-of-states, where the measurement basis can be independent of the specific reference state $\vert S \rangle$. Here, we consider random Pauli measurements.
The set of all Pauli operators can be used to construct a resolution of the identity from the over-complete basis $\{ P_r \vert S \rangle \}$
\begin{align}
    \frac{1}{2^N}\sum_{r} P_r \vert S \rangle \langle S \vert P_r = \mathds{1}.
\end{align}
For this choice, the expression for $D^{A}_R(t)$ becomes
\begin{align}{\label{eqn:DOS_pauli}}
    \langle S \vert A U(t) R \vert S \rangle &= \frac{1}{2^N}\sum_r \langle S \vert A P_r U(t) \vert S \rangle \langle S \vert U^{\dagger}(t) P_r U(t) R \vert S \rangle \\
    &= e^{-i E_S t}\sum_r \langle S \vert A P_r \vert S \rangle C_{P_r, R}(t)
\end{align}
where $E_S$ is the energy of the eigenstate $\vert S \rangle$. This quantity can be efficiently evaluated classically as well. However, note the variance of the estimator for this quantity is generically large, due to the exponential ``shadow norm'' for high-weight operators~\cite{huang_predicting_2020}, see also \eqref{eqn:COR_pauli}. Improving sample complexity for the randomized-measurement version of many-body spectroscopy is an exciting direction for continued research.

\subsubsection{Beyond MPS reference states with doubled Hilbert spaces}

Here, we have proposed a hybrid classical-quantum approach for evaluating $D^{A}(\omega)$, which leverages an efficient classical representation for $\vert R \rangle$ and $\vert S \rangle$, that is suitable for near-term applications.
In some cases,  this requirement can be replaced by a fully quantum measurement, while preserving the sample-efficiency of the procedure, by coherently simulating two systems in parallel and performing entangled measurements between them~\cite{bluvstein_controlling_2021}.

We illustrate such a scheme for estimation of the bare-DOS $D^{\mathds{1}}(\omega)$.
Then, we observe that the sum over Pauli operators in \eqref{eqn:DOS_pauli} can be re-written as a two-copy measurement of the form
\begin{align}
    D_R^{\mathds{1}}(t) = \langle S \vert U(t) R \vert S \rangle &= \frac{1}{2^N} \sum_r \langle R \vert P_r \vert S \rangle \langle S \vert P_r U(t) R \vert S \rangle \\
    &= \frac{1}{2^N} \sum_r \langle R \vert \langle S \vert (P_r \otimes P_r)  (\mathds{1} \otimes U(t)) \vert S \rangle \vert R \rangle
\end{align}
This resulting correlation function has the same form as the two-time correlators measured by Fig.~\ref{fig:spectroscopy_toy_model}a.
Further, all the observables commute $P_r \otimes P_r$, indicating that they can be written as diagonal operators in the correct measurement basis.
This basis corresponds to performing pairwise Bell measurements between the two systems.
To see this, notice that a single pairwise Bell measurement simultaneously measures $\mathds{1}, XX, YY, ZZ$. Therefore, all $4^N$ Pauli-pairs $P_r \otimes P_r$ can be measured in parallel by factorizing into individual pairs between the two systems.
These entangled measurements can be readily implemented in reconfigurable architectures, as demonstrated in Ref.~\cite{bluvstein_quantum_2022}.

\subsection{Thermal expectation values}

Once we have $D^{A}(\omega)$, a simple way to compute thermal expectation values is by reweighting the density of states, and computing the following ratio
\begin{align}\label{eq:finite_temp}
    \langle A \rangle_{\beta} = \frac{\int e^{-\beta \omega} D^{A}(\omega)d\omega }{\int e^{-\beta \omega} D^{\mathds{1}}(\omega)d\omega}
\end{align}
as discussed also in Ref.~\cite{lu_algorithms_2021}.

We further note that this ratio is theoretically robust to uniform spectral broadening, e.g. arising from a finite maximum evolution time.
To illustrate this, we can broaden the delta function in $D^{A}(\omega)$ to be
\begin{align}
    D^{A}_T(\omega) = \sum_n \langle n \vert A \vert n \rangle \delta_T(\omega - \epsilon_n)
\end{align}
where $\delta_T$ is a broadened delta function with width $\sim 1/T$.
When we add the Boltzmann weight $e^{- \beta \omega}$ and integrate over frequencies, the numerator can be rewritten as
\begin{align}
    \int d\omega D^{A}_T(\omega) e^{- \beta \omega} &= \int d\omega \sum_n \langle n \vert A \vert n \rangle \delta_T(\omega - \epsilon_n) e^{- \beta \omega} \\
    &= \sum_n \langle n \vert A \vert n \rangle \int d\omega \delta_T(\omega - \epsilon_n) e^{- \beta \omega} \\
    &= \sum_n \langle n \vert A \vert n \rangle e^{- \beta \epsilon_n} \underbrace{\int d\omega \delta_T(\omega) e^{- \beta \omega}}_{C_T} \\
    &= C_T \sum_n \langle n \vert A \vert n \rangle e^{- \beta \epsilon_n}.
\end{align}
Therefore, we see this is equal to the the numerator when $T \rightarrow \infty$, rescaled by a $T$-dependent constant $C_T$.
The denominator is similarly rescaled by the same constant,
\begin{align}
    \int d\omega D^{\mathds{1}}_T(\omega) e^{- \beta \omega} &= \int d\omega \sum_n \delta_T(\omega - \epsilon_n) e^{- \beta \omega} \\
    &= \sum_n e^{- \beta \omega_n} \int d\omega  \delta_T(\omega) e^{- \beta \omega} \\
    &= C_T \sum_n e^{- \beta \epsilon_n}.
\end{align}
As such, the constant $C_T$ cancels out in the ratio, producing
\begin{align}
    \langle A \rangle_{\beta, T} = \frac{\sum_n \langle n \vert A \vert n \rangle e^{- \beta \epsilon_n}}{\sum_n e^{- \beta \epsilon_n}} = \langle A \rangle_{\beta, T \rightarrow \infty}
\end{align}
which is equal to the ideal value.

\subsubsection{Magnetic Susceptibility}

The magnetic susceptibility for the Heisenberg models studied in Fig.~\ref{fig:spectroscopy_toy_model} and Fig.~\ref{fig:OEC} can be computed straightforwardly by selecting $A = (S^z)^2$.
To understand why this is the case, consider coupling the system to an external magnetic field $\bf{B}$~\cite{alessio_equation--motion_2021}. Define global spin operators $\hat{\bf{S}} = \sum_i \hat{\bf{S_i}}$. Then, the Hamiltonian in the presence of a field becomes
\begin{align}
    H \rightarrow H + \mathbf{B} \cdot \hat{\mathbf{S}}
\end{align}
The density matrix at finite temperature is further given by
\begin{align}
    \rho_{\beta} = \frac{1}{Z}e^{- \beta H}, Z = \tr[e^{- \beta H}].
\end{align}
The magnetization is a vector quantity measured by
\begin{align}
    M^{\mu}(\mathbf{B}, \beta) = \frac{1}{Z} \tr[ S^{\mu} e^{- \beta H}]
\end{align}
and the susceptibily is a response matrix
\begin{align}
    \chi^{\mu}_{\nu} = \frac{\partial}{\partial B^{\nu}} M^{\mu}(\bf{B}, \beta)
\end{align}
which depends on both the temperature $\beta$ and external field $\bf{B}$.
Using the fact that $\bf{S}$ commutes with $H$ in the Heisenberg models, the derivative can be simplified to
\begin{align}
    \chi^{\alpha}_{\beta}(\mathbf{B}, {\beta}) = \beta \tr[\hat{S}^{\alpha} \hat{S}^{\beta} \rho_{\beta}].
\end{align}
In the main text, we focus on $\chi$ at zero-field $\bf{B}=0$, where the rotational symmetry implies the matrix $\chi^{\alpha}_{\beta}$ is proportional to $\delta^{\alpha}_{\beta}$. 
Thus, we focus on one component, and plot
\begin{align}
    \chi(T) = \chi^{z}_{z}(\mathbf{B}=0, {\beta}) = \beta \tr[(\hat{S}^{z})^2 \rho_{\beta}].
\end{align}

\subsubsection{Sampling schemes for low temperatures}

As discussed above, it is important for sample efficiency that the ensemble of perturbations $\mathcal{R}$ prepare probe states that have significant overlap with target eigenstates.
Classical optimization can be used to prepare variational states with larger overlap on, e.g the ground state.

However, to accurately compute finite-temperature properties, while preferentially targetting the low-energy sector, it is crucial that we accurately compute the density of states $D^{A}(\omega)$, without introducing bias by restricting to a smaller subspace.
Importance sampling is a method to simulate uniform sampling over all of Hilbert space, while still preferentially sampling low-energy states.
The key intuition is that we can estimate the low-energy part of $D^{A}(\omega)$ with lower-noise, at the cost of making the high-energy part of $D^{A}(\omega)$ more noisy.
Then, since the contribution to low-temperature properties is dominated by the low-energy part of $D^{A}(\omega)$, this produces an overall less noisy estimate.

\begin{figure}
    \centering
    \includegraphics[width=0.48\textwidth]{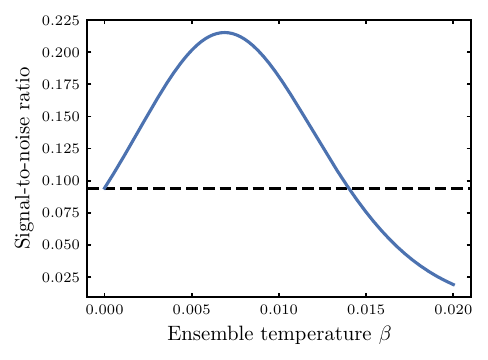}
    \caption{\textbf{Importance sampling to improve SNR}. We consider the signal-to-noise ratio associated with calculating ground-state ($T=0$) properties of the S2H-1b model of the OEC. We consider perturbed states $\vert R \rangle = R \vert {0}^N\rangle$ generated by applying random single-qubit rotations (i.e. random product states), and sample from an ensemble $\mathcal{R}'$ with probabilities proportional to $e^{- \beta_{ensemble} E(R)}$. We find that there is an optimal temperature $\beta_{ensemble}$ which approximately doubles the signal-to-noise ratio compared to a fully random ensemble ($\beta_{ensemble} = 0$). Further increasing $\beta_{ensemble}$ actually causes the signal-to-noise ratio to become worse, since the ground state cannot be expressed as a product state. Initial states supporting some entanglement (e.g. MPS) should further improve the SNR.}
    \label{fig:oec_snr_groundstate}
\end{figure}

Specifically, we consider an importance sampling scheme, where the probability of sampling a state $\vert R \rangle$ from a uniform distribution $\mathcal{R}$ is weighted by the Boltzmann factor $P_{\mathcal{R}'}(R) \propto e^{-\beta E(R)}$ where $E(R) = \langle R \vert H \vert R \rangle$ is a clasically computed energy.
Then, we can replace the average over $\mathcal{R}$ with a weighted average over $\mathcal{R'}$.
\begin{align}
    \mathbb{E}_{R \sim \mathcal{R}} 2^N \rightarrow \mathbb{E}_{R \sim \mathcal{R}'} \frac{2^N}{P_{\mathcal{R}'}(R)}
\end{align}
Substitutions of this form can reduce the total sample complexity, as we demonstrate for the random product state case in Extended data Fig.~\ref{fig:oec_snr_groundstate}.

\subsection{Efficient evaluation of spin-sector projectors}

Our efficient estimator \eqref{eqn:ORA_estimator} of the operator-resolved density of states involves evaluating expectation values of the form $\langle R \vert A \vert b \rangle$ where $b$ is the measured ($X$-basis) product state, $\vert R \rangle$ is the initially prepared probe state, and $A$ is the operator to measure.
As interesting choices for $A$, we utilize the projector onto a total $S^z$ sector, and the projector onto total-spin sectors $P_s$.

To evaluate these overlaps efficiently, we utilize the formalism of 1D tensor networks.
In particular, write the projector onto an $S^z$ sector with $m$ spin-flips as a matrix-product operator 
\begin{align}
P(N=N, S^z = \frac{N}{2} - m) = \langle m \vert_a \prod_{i=1}^{N} \left(\pi^{0}_i \mathds{1}_a + \pi^{1}_i \sigma^+_a \right) \vert 0 \rangle_a
\end{align}
where in the term $\pi^1_i \sigma^+_a$, the first index acts on the physical site $i$, and the second term acts on the ancillary mode, which starts at $\vert 0 \rangle$ on the right and at the end is projected onto $\langle m \vert$.
Intuitively, the ancilla mode counts the number of qubits in the state $\vert {1}\rangle$, and the ket $\langle m \vert$ only picks out strings with exactly $m$ down spins.
Thus, this forms the projector onto all states with exactly $m$ down spin, or $S^z = \frac{N}{2}-m$.
This MPO has bond dimension $\Theta(N)$.
Therefore, overlaps of the form $\langle R \vert P_{S^z} \vert b \rangle$, where $\vert R \rangle$ and $\vert b \rangle$ are both finite-bond dimension MPS, can be efficiently evaluated in $O(N^2)$ time via tensor contraction.

Next, we generalize the projector construction to sectors with well-defined total-spin and $z$-component, $S$ and $S^z$ respectively.
The key insight underlying the construction is to utilize three ancilla modes, which track the total $S^z$ of the bra and ket, as well as their Hamming distance.
\begin{align}
P(N=N, S=\frac{N}{2} - s, S^z = \frac{N}{2} - m) &= \langle N,s,m \vert_a \times \nonumber\\
&\prod_{i=1}^{N} \left( 
\vert {0} \rangle_i \langle {0} \vert_i (\mathds{1 1 1})_a 
+ \vert {0} \rangle_i \langle {1} \vert_i (\mathds{1} \sigma^+ \sigma^+)_a
+ \vert {1} \rangle_i \langle {0} \vert_i (\sigma^+  \mathds{1}\sigma^+)_a 
+ \vert {1} \rangle_i \langle {1} \vert_i (\sigma^+ \sigma^+ \mathds{1})_a 
\right) \nonumber \\
&\times \vert 000 \rangle_a
\end{align}
The bond-dimension of this MPO scales as $O(N^3)$ in general.
To understand this expression, notice that we can always write the matrix elements of the projector as $P = \sum_{s,s'} P_{s,s'} \vert s \rangle \langle s' \vert$  in the $z$-basis, where the coefficient $P_{s,s'} = \langle s \vert P \vert s' \rangle$.
When the MPO for $P$ is contracted with the two bistrings $\langle s \vert \cdot \vert s' \rangle$, the expression reduces to
\begin{align}
    P_{s,s'} = \langle N,s,m \vert w(s) w(s') w(s-s') \rangle_a
\end{align}
where $w(s)$ measures the number of down-spins (weight) of each configuration $s$.
As in the $S^z$ case, for this expression to be non-zero the number of down-spins in both $s$ and $s'$ must equal $m$, which imposes the condition $w(s)=w(s')=m$.
However, in general, the projector also contains off-diagonal terms, where the Hamming distance $w(s-s')$ between the bra and ket state can be non-zero.

A simple example is the projector onto the triplet state, 
\begin{align}
    P(N=2,S=1, S^z=0) = \frac{1}{2}\left(\vert {0}{1} \rangle\langle {0}{1} \vert + \vert {0}{1} \rangle\langle {1}{0} \vert + \vert {1}{0} \rangle\langle {0}{1} \vert + \vert {1}{0} \rangle\langle {1}{0} \vert \right)
\end{align}
For this example, we see the termination vector should take the form
\begin{align}
    \vert 2,0,1 \rangle = \frac{1}{2} \left( \vert 1 1 0 \rangle + \vert 1 1 2 \rangle \right)
\end{align}
Note that this object is a virtual state, not a physical state, and hence does not need to be normalized.
To simplify notation, we will in general write each termination vector as
\begin{align}
    \vert N, s, m \rangle_a = \sum_{d=0}^{m} P^{N}_{s,m,d} \vert (m) (m) (2d) \rangle_a
\end{align}
where $P_{s,m,d}^{N}$ encodes all of the relevant weights.
We note that, in this expression, we use the fact that the Hamming distance between two configurations with the same $S^z$ is always an even number $2d$, and is can be no greater than twice the number of spin-flips $m$. As such, we can interpret $d$ as the number of exchanges needed to connect the two-strings $s$ and $s'$.

We present an efficient ($\mathrm{poly}(N)$ time) recursive construction for the coefficients $P_{s,m,d}^{N}$

To illustrate the idea, let us start with the simplest projector, onto $S = \frac{N}{2}$ and $S^z = \frac{N}{2}$, or $s=0,m=0$.
The only state which contributes is the all-up state, so $P^{N}_{0,0,0} = 1$.
The projectors onto all other symmetric subspaces, with $S=\frac{N}{2}, S^z=\frac{N}{2}-m$ can be computed using a ladder-like construction.
For one spin-flip ($m=1$), the projector can be written as
\begin{align}
    P(N=N, S=\frac{N}{2},S^z=\frac{N}{2}-1) = \frac{1}{\mathcal{N}} \left(\sum_i \sigma_i^+ \right) P(S=\frac{N}{2},S^z=\frac{N}{2}-1) \left(\sum_i \sigma_i^- \right).
\end{align}

By plugging in representative configurations with a single down-spin, we can compute the relevant coefficients to be 
\begin{align}
    P^{N}_{0,1,0} = 1 / \mathcal{N} & & P^{N}_{0,1,1} = 1 / \mathcal{N}
\end{align}
To compute the normalization factor $\mathcal{N}$, we use the fact that $P^2{=}P$.
In particular, we can again plug in representative configurations to match
\begin{align}
    \langle s \vert P \vert s' \rangle &= \langle s \vert P^2 \vert s' \rangle \\
    &= \sum_{s''} \langle s \vert P \vert s'' \rangle \langle s'' \vert P \vert s' \rangle
\end{align}
Taking $\vert s \rangle = \vert s'\rangle = \vert {1}{0}^{N-1} \rangle$ as the representative states, we could explicitly sum over the $N$ relevant intermediate states.
However, to simplify the calculation, we note the intermediate states naturally split into two groups.
The first is distance-0 from $s$ and $s'$, i.e. $\vert s'' \rangle = \vert s \rangle$, which contributes $(P_{0,1,0}^{N})^2$. 
The second group is distance-2 from $s$ and $s'$, such as $\vert s'' \rangle = \vert {0}{1}{0}^{N-2} \rangle$. This contributes $(P^{N}_{0,1,1})^2$, and there are $N-1$ such contributions.
Therefore, we get the equality
\begin{align}
    P_{0,1,0}^{N} &= (P_{0,1,0}^{N})^2 + (N-1) (P_{0,1,1,}^{N})^2 \\
    1/\mathcal{N} &= N / \mathcal{N}^2 \\
    \mathcal{N} &= N.
\end{align}
We could similarly select $\vert s \rangle$ and $\vert s' \rangle$ so the left-hand side is the term $P_{0,1,1}^{N}$ corresponding to Hamming distance 2.
There, a similar calculation over intermediate paths leads to the constraint
\begin{align}
    P_{0,1,1}^{N} &= 2(P_{0,1,0}^{N} P_{0,1,1}^{N}) + (N-2)(P_{0,1,1}^{N})^2 \\
    \mathcal{N} &= N
\end{align}
resulting in the same normalization factor.
Indeed, the condition $P = P^2$ implies the computed value for $\mathcal{N}$ must be the same for all non-zero coefficients appearing in the left-hand side.
The key intuition underlying these calculations, is that the valid paths can be efficiently partitioned into groups, and the coefficients can be efficiently computed via combinatorics.

The same intuition can be generalized to a recursive algorithm for computing $P_{s,m,d}$ from $P_{s,m-1,d'}$, which allows us to compute the full ladder of states with $m=0,...,N$ from the root cases. 
First, we compute the unnormalized coefficients using the formula
\begin{align}
    P'_{s,m,d} = d^2 P_{s,m-1,d-1} + (m-d)(2d+1) P_{s,m-1,d} + (m-d)(m-d-1) P_{s,m-1,d+1}
\end{align}
Where we assume $P_{s,m,d}=0$ when $d > m$.
To understand the coefficient associated with each term, we consider a pair of representative states $\vert s \rangle = \vert \1^{m-d} \1^d \0^d \0^{N-m-d} \rangle$ and $\vert s' \rangle = \vert \1^{m-d} 0^d 1^d 0^{N-m-d} \rangle$, which have $m$ down-spins and Hamming distance $2d$. For convenience, we label the four groups of size $m-d$, $d$, $d$, and $N-m-d$ as $A$,$B$,$C$, and $D$ respectively. The formula comes from the following case-by-case breakdown:
\begin{enumerate}
    \item The term $\vert s \rangle \langle s' \vert$ can come from a pair with smaller Hamming distance $2(d-1)$ by flipping one spin from $B$ in $s$ and from $C$ in $s'$. There are $d^2$ such choices.
    \item The term $\vert s \rangle \langle s' \vert$ can come from a pair with the same Hamming distance $2d$, by flipping one spin from $A$ in $s$ and one spin from $C$ in $s'$, vice versa, or by flipping the same spin from $A$ for $s$ and $s'$. There are $2(m-d)d + (m-d) = (m-d)(2d+1)$ such choices.
    \item The term $\vert s \rangle \langle s' \vert$ can come from a pair with larger Hamming distance $2(d+1)$ by flipping two different spins from $A$ in $s$ and $s'$. There are $(m-d)(m-d-1)$ of these pairs.
\end{enumerate}

Then, we need to compute the normalization factor.
In order to do this, consider the same pair of representative states, and enumerate over all intermediate states $s''$.
In particular, we can efficiently label intermediate states by counting the number of down-spins in each of the four groups, $m_A, m_B, m_C, m_D$, where the total number of down-spins satisfies $m = m_A + m_B + m_C + m_D$.
For each intermediate configuration, the distance from $s$ to $s''$ and $s'$ to $s''$ is given by
\begin{align}
    d(s,s'') =  (m-d - m_A) + (d - m_B) + m_C + m_D \\
    d(s',s'') =  (m-d - m_A) + m_B + (d - m_C) + m_D
\end{align}
The number of intermediate states $s''$ associated with each label is
\begin{align}
    Z(m_A, m_B, m_C, m_D) = {m-d \choose m_A} {d \choose m_B} {d \choose m_C} {N-m-d \choose m_D}.
\end{align}
Taken together, by enumerating over valid assignments for the number of down-spins in each group, we can compute the normalization factor
\begin{align}
    \mathcal{N} = \frac{\sum_{m_A+m_B+m_C+m_D = m} Z(m_A, m_B, m_C, m_D) P'_{s,m,d(s,s'')} P'_{s,m,d(s',s'')}}{P'_{s,m,d(s,s')}}
\end{align}
and the normalized projector coefficients $P_{s,m,d} = P' / \mathcal{N}$.

Next, we discuss projectors onto states where the total-spin is smaller then $N/2$.
To illustrate this, we start by computing $P(S=\frac{N}{2}-1,S^z=\frac{N}{2}-1)$.
To do this, we see that there are only two total-spin sectors which contribute to $S^z=\frac{N}{2}-1$, $s=0$ and $s=1$.
Therefore, we know that
\begin{align}
    P(N=N, S=\frac{N}{2}-1,S^z=\frac{N}{2}-1) = P(N=N, S^z=\frac{N}{2}-1) - P(N=N, S=\frac{N}{2},S^z=\frac{N}{2}-1)
\end{align}
The termination vector associated with $S^z = \frac{N}{2}-m$ sector is $\vert (m)(m)(0) \rangle$.
Therefore, the coefficients of the projector can be written as $P_{1,1,d}^{N} = \delta_{d,0} - P_{0,1,d}^{N}$.
As an example, consider the $N=2$ singlet state, for which this formula implies $P_{1,1,0}^{2} = 1/2$ and $P_{1,1,2}^{2}=-1/2$.
This leads to the familiar expression
\begin{align}
    P(N=1,S=0, S^z=0) = \frac{1}{2}\left( \vert \0 \1 \rangle \langle \0 \1 \vert - \vert \1 \0 \rangle \langle \0 \1 \vert - \vert \0 \1 \rangle \langle \1 \0 \vert + \vert \1 \0 \rangle \langle \1 \0 \vert \right).
\end{align}

More generally, to construct each of the terms in the root projector with $s=m$, we orthogonalize w.r.t all higher-spin sectors.
\begin{align}
    P_{m,m,d} = \delta_{d,0} - \sum_{0 \leq s < m} P_{s,m,d}
\end{align}
Here, each term with $s<m$ can be computed from root terms where $s=m$ using the recursive ladder approach described above.

\end{document}